\documentclass{aa} 

\usepackage{graphicx}
\usepackage{txfonts}
\usepackage{xcolor}

\begin{document}

   \title{The CARMENES search for exoplanets around M dwarfs}

   \subtitle{Rubidium abundances in nearby cool stars}
   
   \titlerunning{Rubidium abundances in CARMENES M dwarfs}
   
   \authorrunning{C. Abia et al.}

   \author{C.\,~Abia\inst{1},
        H.\,M.~Tabernero\inst{2,4},
        S.\,A.~Korotin\inst{3},
        D.~Montes\inst{4},
        E.~Marfil\inst{4},
        J.\,A.~Caballero\inst{5},
        O.~Straniero\inst{6},
        N.~Prantzos\inst{7},
        I.~Ribas\inst{8,9},
        A.~Reiners\inst{10},
        A.~Quirrenbach\inst{11},
        P.\,J.~Amado\inst{12},
        V.\,J.\,S.~B\'ejar\inst{13,14},
        M.~Cort\'es-Contreras\inst{8},
        S.~Dreizler\inst{9},
        Th.~Henning\inst{15},
        S.\,V.~Jeffers\inst{10},
        A.~Kaminski\inst{11},
        M.~K\"urster\inst{15},
        M.~Lafarga\inst{8,9},
        \'A.~L\'opez-Gallifa\inst{4},
        J.\,C.~Morales\inst{8,9},
        E.~Nagel\inst{15},
        V.\,M.~Passegger\inst{17,18},
        S.~Pedraz\inst{19},
        C.~Rodr\'iguez~L\'opez\inst{12},
        A.~Schweitzer\inst{17},
        M.~Zechmeister\inst{10}
        }

   \institute{Departamento de F\'\i sica Te\'orica y del Cosmos, Universidad de Granada, E-18071 Granada, Spain\\
    \email{cabia@ugr.es}
    \and
    Instituto de Astrof{\'i}sica e Ci{\^e}ncias do Espa\c{c}o, Universidade do Porto, CAUP, Rua das Estrelas, 4150-762 Porto, Portugal
    \and 
    Crimean Astrophysical Observatory, Nauchny 298409, Republic of Crimea 
    \and
    Departamento de F{\'i}sica de la Tierra y Astrof{\'i}sica \& IPARCOS-UCM (Instituto de F\'{i}sica de Part\'{i}culas y del Cosmos de la UCM),
    Facultad de Ciencias F{\'i}sicas, Universidad Complutense de Madrid, E-28040 Madrid, Spain
    \and 
    Centro de Astrobiolog\'{i}a (CSIC-INTA), ESAC, Camino bajo del castillo s/n, E-28691 Villanueva de la Ca\~{n}ada, Madrid, Spain
    \and
    Istituto Nazionale di Astrofisica - Osservatorio Astronomico d’Abruzzo, Via Maggini snc, I-64100,
              Teramo, Italy
    \and
    Institut d’Astrophysique de Paris, UMR7095 CNRS, Univ. P. \& M. Curie, 98bis Bd. Arago, F-75104 Paris, France
    \and
    Institut de Ci\`{e}ncies de l’Espai (CSIC-IEEC), Campus UAB, c/ de Can Magrans s/n, E-08193 Bellaterra, Barcelona, Spain
    \and
    Institut d’Estudis Espacials de Catalunya (IEEC), E-08034 Barcelona, Spain
    \and
    Institut f\"{u}r Astrophysik, Georg-August-Universit\"{a}t, Friedrich-Hund-Platz 1, D-37077 G\"{o}ttingen, Germany
    \and
    Landessternwarte, Zentrum f\"{u}r Astronomie der Universt\"{a}t Heidelberg, K\"{o}nigstuhl 12, D-69117 Heidelberg, Germany
    \and
    Instituto de Astrof\'{i}sica de Andaluc\'{i}a (IAA-CSIC), Glorieta de la Astronom\'{i}a s/n, E-18008 Granada, Spain
    \and
    Instituto de Astrof\'{i}sica de Canarias, c/ V\'{i}a L\'{a}ctea s/n, E-38205 La Laguna, Tenerife, Spain
    \and
    Departamento de Astrof\'{i}sica, Universidad de La Laguna, E-38206 La Laguna, Tenerife, Spain
    \and
    Th\"{u}ringer Landessternwarte Tautenburg, Sternwarte 5, D-07778 Tautenburg, Germany
    \and
    Max-Planck-Institut f\"{u}r Astronomie, K\"{o}nigstuhl 17, D-69117 Heidelberg, Germany
    \and
    Hamburger Sternwarte, Gojenbergsweg 112, D-21029 Hamburg, Germany
    \and
    Homer L. Dodge Department of Physics and Astronomy, University of Oklahoma, 440 West Brooks Street, Norman, OK-73019 Oklahoma, United States of America
    \and
    Centro Astron\'{o}mico Hispano-Alem\'{a}n (CSIC-MPG), Observatorio de Calar Alto, Sierra de los Filabres, E-04550 G\'{e}rgal, Spain
    }
             
   \date{Received dd July 2020 / Accepted dd Month 2020}

   \abstract{Due to their ubiquity and very long main-sequence lifetimes, abundance determinations in M dwarfs
     provide a powerful and alternative tool to GK dwarfs to study the formation and chemical enrichment history of our Galaxy. 
In this study, abundances of the neutron-capture elements Rb, Sr, and Zr are derived, for the first time, in a sample of nearby M dwarfs. 
We focus on stars in the metallicity range $-0.5\lesssim$ [Fe/H] $\lesssim +0.3$, an interval poorly explored for Rb abundances in previous analyses. 
To do this we use high-resolution, high-signal-to-noise-ratio, optical and near-infrared spectra of 57 M dwarfs observed with CARMENES. 
The resulting [Sr/Fe] and [Zr/Fe] ratios for most M dwarfs are almost constant at about the solar value, and are identical to those found in
GK dwarfs of the same metallicity. 
However, for Rb we find systematic underabundances ([Rb/Fe]$<0.0$) by a factor two on average. Furthermore, a tendency is found for Rb---but
not for other heavy elements (Sr, Zr)---to
increase with increasing metallicity such that [Rb/Fe]$\gtrsim 0.0$ is attained at metallicities higher than solar. 
These are surprising results, never seen for any other heavy element, and are difficult to understand within the formulation of the s- and
r-processes, both contributing sources to
the Galactic Rb abundance. 
We discuss the reliability of these findings for Rb in terms of non-LTE (local thermodynamic equilibrium) effects, stellar activity,
or an anomalous Rb abundance in the Solar System, but no explanation is found. We then interpret the full observed [Rb/Fe] versus [Fe/H] trend
within the framework of theoretical predictions from state-of-the-art chemical evolution models for heavy elements, but a simple interpretation
is not found either. In particular, the possible secondary behaviour of the [Rb/Fe] ratio at super-solar metallicities would require a much larger
production of Rb than currently predicted in AGB stars through the s-process without overproducing Sr and Zr. }
 

   
   \maketitle
\section{Introduction}
   
Analysis of the rubidium abundance in the Solar System shows that the neutron capture s- and r-processes are about equally responsible for the
synthesis of this element \citep[e.g.][]{sne08,pra20}. 
Rubidium is present in two isotopic forms: $^{85}$Rb, which is stable, and $^{87}$Rb, which with a half-life of $5\times10^{10}$\,yr may be considered
stable from an astrophysical point of view. 
Astronomical detection of Rb relies on the \ion{Rb}{I} lines, which do not permit measurement of the relative isotopic Rb abundances from stellar spectra. 
The main s-process, which manufactures elements with atomic mass in the range $90 < A < 209$, is identified with the He-burning shell in low- and
intermediate-mass stars (LIMS; $M \lesssim \,8$ M$_\odot$) during the asymptotic giant branch (AGB)
phase \citep[e.g.][]{bus99,kap11},
where neutrons are mainly provided by the $^{13}$C$(\alpha,$n$)^{16}$O reaction. 
The weak s-process, responsible for a major contribution to the s-process nuclides up to $A\sim 90$, has been recognised as the result of neutron
capture synthesis mainly during core He- and shell C-burning phases of massive stars ($M \gtrsim 10$ M$_\odot$; \citealt{arn85,pra90,rai91,pig10,lim18})
with the reaction $^{22}$Ne$(\alpha,$n$)^{25}$Mg being the major neutron source.

The situation concerning the astrophysical site of the r-process is still far from clear. 
After more than 50 years of research into its astrophysical origin(s), the identification of a fully convincing site remains elusive. 
In the recent past, the r-process was usually associated to the explosion of massive stars, but nowadays the neutron-star merging scenario is
gaining support, bolstered by the recent joint detection of electromagnetic and gravitational signals from the $\gamma$-ray burst
\object{GW~170817}/GRB~170817A \citep[see][and references therein]{pia17}. 
Nevertheless, the behaviour of the abundance ratios with respect to Fe of genuine r-process elements, such as Eu, implies that core
collapse supernova may also play an important role in r-process nucleosynthesis. 
In any case, to date, no numerical simulation in the proposed scenarios is able to fully reproduce the observed distribution of
the r-process elemental and isotopic abundances in the Solar System. 
Excellent reviews on this topic can be found in \citet{thi17}, \citet{cow19}, and \citet{arn20}.

There is substantial observational evidence showing that Rb is produced through the main s-process in AGB stars. Abundance analyses of
many Galactic and extragalactic
AGB stars both of C- and O-rich types show
considerable Rb enhancements, namely [Rb/Fe] $>0.0$\footnote{Here we follow the standard abundance notation,
  [X/H] $= \log{({\rm X/H})_\star} - \log{({\rm X/H})_\odot}$, where X/H is the abundance by number of the element X,
  and $\log \epsilon(X) \equiv \log{({\rm X/H})} + 12$.}
\citep{lam95,abi98,abi01,gar06,del06,gar09,per17}. 
Generally, these enhancements can be explained by the current nucleosynthesis models of AGB stars \citep{gallino1998,cri11,van12,kar14,cri15}. 
Two are the possible neutron sources in AGB stars: the  $^{13}$C$(\alpha,$n$)^{16}$O reaction, which is the main neutron source in low-mass AGB stars,
and the $^{22}$Ne$(\alpha, $n$)^{25}$Mg reaction, which becomes important in massive AGBs. 
Theoretical models predict [Rb/Sr,Y,Zr] ratios lower or larger than solar depending on whether the $^{13}$C$(\alpha,$n$)^{16}$O  or
the $^{22}$Ne$(\alpha, $n$)^{25}$Mg reaction is the main neutron source, respectively. The difference is in the neutron density, which
is rather low for the $^{13}$C$(\alpha,$n$)^{16}$O  reaction (N$_{\rm n}\sim 10^7$\,cm$^{-3}$), but much larger in the case of the $^{22}$Ne$(\alpha, $n$)^{25}$Mg 
reaction (N$_{\rm n}> 10^{11}$\,cm$^{-3}$). 
Abundance ratios [Rb/Sr,Y,Zr] $<0.0$ are indeed found in most carbon-rich AGB stars, which indicates their low-mass AGB origin ($M \sim 1.3-3.0$\,M$_\odot$),
while positive ratios are found in very luminous O-rich AGB stars with masses $M \gtrsim 4$\,M$_\odot$, in which the $^{22}$Ne neutron source is
dominant \citep{lam95,abi01,per17}.

The evolution of the Galaxy's s- and r-process products on the other hand is directly observed from the stellar abundances of elements that are predominantly
attributable to either the s- or the r-process \citep[see e.g.][]{jof19}. 
Traditional tracers include Sr, Ba, and La for the s-process and Eu as the genuine representative element of the r-process\footnote{\citealt{pra20} derived
  an s-process contribution in the Solar System abundances of 91\,\%, 88\,\%, and 80\,\% for
Sr, Ba, and La, respectively, while the r-process contribution for Eu is 95\,\%.}. 
Due to the nearly equal s- and r-process production of Rb, the study of the evolution of the abundance of this element with metallicity is particularly
suitable for elucidating the timescales on which both neutron capture processes contribute to the chemical history of the Galaxy.
However, the unfavourable electronic structure of Rb provides only a few lines in stellar spectra. 
The most used lines are the  \ion{Rb}{I} resonance lines at $\lambda\lambda$7800.3 and 7947.6\,{\AA}. 
Due to the low cosmic abundance of Rb, these lines are usually weak, and in G dwarfs they are also heavily blended. 
Because of their weakness, the blend issue is aggravated in stars with near solar metallicity (or higher). 
To date, only two Rb abundance studies in these stars exist, namely those of \citet{gra94} and \citet{tom99}. 
These authors derived Rb abundances in a sample of metal-poor disc and halo G dwarfs and concluded that the [Rb/Fe] versus [Fe/H] relationship behaves at
low metallicity ([Fe/H] $<-1$) as does [Eu/Fe], that is, showing an approximately constant [Rb/Fe] ratio as a typical r-process element. 
However, the behaviour at higher metallicities ([Fe/H] $\gtrsim -0.5$) is poorly understood.
It is precisely at those metallicities that the s-process contribution to the Galactic Rb abundance is expected to peak \citep[e.g.][]{bus99,cri15,bis17,pra18}. 
Detection via the \ion{Rb}{I} resonance lines at the expected low Rb abundances is more favourable for cool dwarfs and giants. Due to the
lower effective temperature
and higher gravity of M dwarfs, the Rb resonance lines show a moderate intensity in these stars and can be relatively easily identified in their spectra for a
wide range of metallicities, in particular in stars with metallicity close to solar using high spectral resolution. 
However, there is a caveat to using these lines: although prominent, they are contaminated with the presence of a plethora of TiO and VO absorption lines
and as a consequence the spectral continuum is usually difficult to locate.

In this study we derive Rb abundances from high-resolution spectra in a sample of nearby, bright, single M dwarfs with metallicities close to solar.
Simultaneously, we also {determined} their Sr and Zr abundances, which are two elements with a predominantly main s-process origin.
For that, we use the high signal-to-noise template spectra of 57 M dwarfs observed under the CARMENES exoplanet survey guaranteed time
observations \citep[GTOs;][]{qui18,Rei18}.
The derived abundance ratios are compared with state-of-the-art chemical evolution model predictions  for the solar neighbourhood. 
Our aim is to better identify the onset of the significant s-process contribution in the chemical history of the Galaxy and put constraints on the
role played by the s- and r-processes in the Galactic Rb budget. 
The structure of this paper is as follows: the observational material and analysis are presented in Section~\ref{sample}, where the data acquisition
and reduction procedures are briefly described; we also discuss the atmospheric parameters used in this study, the line lists, and the derivation of
the abundances from the spectra, together with an evaluation of the observational and analysis uncertainties. 
In Section~\ref{discussion} the main results are discussed and compared with the current nucleosynthesis models through a chemical evolution model
for the solar neighbourhood. 
Finally, Section~\ref{summary} summarises the main conclusions of this study.

\section{Stellar sample and analysis}
\label{sample}
\subsection{Targets, observations, and reduction}

Installed at the 3.5~m telescope at Calar Alto Observatory in Almer\'ia, Spain, CARMENES is a new-generation spectrograph
designed to detect Earth-mass planets around M dwarfs by means of the radial-velocity technique \citep{qui14,qui18}. 
CARMENES consists of two optical and near-infrared (NIR) channels that extend over 5200--9600\,{\AA} in the visual (VIS) and
9600--17100\,{\AA} in the NIR regions with resolving powers of $R$~$=$~$94,600$ and $R$~$=$~$80,400$, respectively. 
The CARMENES exoplanet radial-velocity survey comprises about 350 M dwarfs that cover the full M sequence from M0.0\,V to M9.0\,V
spectral types \citep[and a couple of K7\,V stars;][]{Rei18}. 
The CARMENES GTO sample contains a set of carefully selected M dwarfs in the immediate solar neighbourhood that are the brightest
stars in each spectral subtype without physical companions at less than 5\,arcsec \citep{Cab16a}. 
Most of the target stars should have close to solar composition, although there are some known exceptions \citep{Alo15,pass18,pass19}.
Almost all the stars studied here are located within a few tens of parsecs around the Sun.
However, there are members in all the different Galactic kinematic populations except for the halo, that is, the young thin,
thin-to-thick transition, and thick discs \citep{cor16}, which ensures a broad investigated metallicity range.

All investigated M dwarfs were spectroscopically observed with CARMENES at a number of epochs since January 2016 \citep{Rei18}. The median number
of observations per star in the VIS channel is 24.
Their spectra were reduced according to the standard CARMENES GTO data flow and thus were processed and wavelength calibrated with the {\tt caracal}
pipeline \citep{Cab16b}, which computes a single co-added template spectrum per star and instrument channel \citep[see also][]{Zech18}.
The wavelength calibration is provided by a combination of hollow cathode lamps and two temperature- and pressure-stabilised Fabry-P\'{e}rot
units that
eventually provide median uncertainties of the level of 1\,m\,s$^{-1}$ in the VIS channel \citep{tri18,zech19}, which covers the investigated
\ion{Rb}{i}, \ion{Sr}{ii},
and \ion{Zr}{i} lines.
In addition, we added a step in the data flow, namely correcting the individual spectra from telluric correction before merging them in the template spectra.
This correction provided a further improvement in the S/N of the eventually used spectra, up to about 300.
Further details on the removal of the telluric features can be found in \citet{nag20xx}.

We selected a sample of 57 M dwarfs from the CARMENES GTO programme in a wide metallicity range ($-0.5\leq$ [Fe/H] $\leq +0.3$), imposing $T_{\rm eff} >3400$\,K
to minimise the impact of molecular absorption, according to the values reported by \citet{pass19}.
This metallicity interval was poorly explored for Rb abundances in the literature.

\subsection{Stellar parameter refinement}

A number of studies focused on improving the precision of M  dwarf stellar parameters, namely $T_{\rm eff}$, surface gravity ($\log{g}$), and metallicity,
can be found in the recent literature. 
For instance, different ways to determine the first two parameters were discussed by \citet{vey17}, \citet{sch19}, and \citet{sou20}, while \citet{bir20}
discussed the derivation of the metallicity. 
In contrast to FGK dwarfs, the spectroscopic method to derive these parameters is, in general, not recommended for M dwarfs because of their much
more complex spectra. 
Due to their relatively low effective temperatures ($T_{\rm eff}\lesssim 4000$\,K), the spectra of M dwarfs show a plethora of molecular absorption
lines that leave almost no atomic line free of blends. 
Furthermore, the C/O ratio is revealed as an additional relevant parameter in the analysis. 
The envelopes of M dwarfs are not expected to be affected by the nuclear burning in the stellar interior, and therefore they should show a C/O
identical to that of the interstellar medium (ISM) from which they were born, that is, lower than unity. 
This means that almost all the available carbon in the atmospheres of M dwarfs will be locked into forming CO, while the remaining available
oxygen would form metallic oxides, water, and other, less abundant O-bearing molecules. In fact, the VIS spectra of M dwarfs are dominated by
TiO and VO absorption bands (and some CaH; \citealt{kir91,Alo15}),
whereas  CO, OH, and H$_2$O prevail in the NIR region. 
Therefore, to determine the stellar parameters of M dwarfs, a full spectral synthesis analysis is recommended instead of studying
individual spectral lines. 
To do this, an estimation of C/O is vital as the intensity of the above-mentioned O-bearing molecules mainly depends on this ratio. 
As a consequence, this combined molecular absorption systematically depresses the spectral continuum, which is usually not well
defined \citep[e.g.][]{tsu14}. 

In a first step we adopted the stellar parameters derived by \citet{pass19}, 
who fitted an updated version grid of  the PHOENIX synthetic spectra presented by \citep{hus13} to the selected high-resolution
CARMENES spectra in the VIS and NIR spectral ranges using a $\chi^2$ method. 
We refer to their work for further details on the specific method followed,
for a detailed discussion on the evaluation of the uncertainties, and for comparison with the most recent literature. 
In general, the agreement with other studies was good within the uncertainties, although for some stars with metallicity higher
than solar, \citet{pass19} derived systematically higher metallicities. 

With the stellar parameters estimated as mentioned above, a model atmosphere was interpolated within the grid of MARCS atmosphere
models \citep{gus08} for the specific $T_{\rm eff}$, $\log{g}$, and [M/H] (i.e. average metallicity) derived for each star.  
A microturbulence parameter of $\xi =$ 1\,km\,s$^{-1}$ was adopted for all the stars. 
Certainly, other microturbulence parameters might be possible for M dwarfs, but the severe blending prevented us for deriving this
parameter spectroscopically. 
However, uncertainties of 0.5\,km\,s$^{-1}$ in the microturbulent velocity result in very small changes in the derived abundances (see below). 
The atomic line lists were taken from the VALD3 database adopting the corrections performed within the {\em Gaia}-ESO survey \citep{hei15b}
in the VIS wavelength range, while in the NIR range we used the atomic line list from the APOGEE survey (\citealt{has16,cun17,maj17};
Smith et al. in~prep.). We refer to these surveys for a detailed description of these line lists.
Molecular line lists were provided by B.~Plez\footnote{These molecular line lists are publicly available
at \url{https://nextcloud.lupm.in2p3.fr/s/r8pXijD39YLzw5T}, where detailed bibliographic sources can be also found.}
and they include several C- and O-bearing molecules (CO, CH, CN, C$_2$, HCN, TiO, VO, H$_2$O) and a few metallic hydrides (FeH, MgH, CaH).  
TiO and H$_2$O determine, in particular, the continuum level at VIS and NIR wavelengths, respectively.

\begin{table}
\caption{Spectroscopic data of the used lines and hyperfine structure of \ion{Rb}{I}.}
\label{rb_lines}
\centering
\begin{tabular}{l cc } %
\hline
\hline
\noalign{\smallskip}
Element/ & $\lambda$ & $\log{gf}$ \\
isotope & (\AA) &     \\
\noalign{\smallskip}
\hline
\noalign{\smallskip}
$^{87}$Rb & &  \\ 
 &7800.183 & --0.663 \\
 &7800.186 & --0.663 \\
 &7800.188 & --1.061 \\
 &7800.317 & --0.216 \\
 &7800.322 & --0.663 \\
 &7800.325 & --1.362 \\
& 7947.507 & --0.669 \\
& 7947.524 & --1.368 \\
& 7947.651 & --0.669 \\
&7947.668  & --0.669 \\
$^{85}$Rb & &  \\
 &7800.233 & --0.744 \\
 &7800.234 & --0.647 \\
 &7800.235 & --0.760 \\
 &7800.292 & --0.282 \\
 &7800.295 & --0.647 \\
 &7800.296 & --1.191 \\
 &7947.563 & --0.653 \\
 &7947.570 & --1.197 \\
 &7947.626 & --0.750 \\
 &7947.634 & --0.653 \\
 Sr {\sc ii} & &  \\
 &10327.311 & --0.353 \\
 Zr {\sc i} & &  \\
 &8063.090  & --1.620 \\
 &8070.080  & --0.790 \\
\noalign{\smallskip}
\hline
\end{tabular}
\end{table}

Oscillator strengths for the \ion{Rb}{I} resonance lines at $\lambda\lambda$7800 and 7947\,\AA\ are known with good
accuracy \citep{Morton00}, but they have multiple components due to the hyperfine structure of the atomic levels. 
The individual  hyperfine structure components and their respective oscillator strengths are given in Table~\ref{rb_lines}. 
We adopted the meteoritic $^{85}$Rb/$^{87}$Rb $=2.43$ ratio \citep{lod19}. Unfortunately, the isotopic splitting is tiny and does
not  allow the derivation of the $^{85}$Rb/$^{87}$Rb ratio from our spectra. Concerning Sr, this element has only a few lines useful
for abundance determinations. 
In fact, the available strontium lines (usually \ion{Sr}{II}) in the blue spectral region of cool stars with near-solar
metallicity are usually strongly blended.
An alternative are the three \ion{Sr}{II} lines in the NIR at $\lambda\lambda$10036, 10327, and 10914\,\AA, which are almost
free of blends and telluric contamination. 
However, as shown in \citet{Belyakova97}, \citet{Andrievsky11}, and \citet{Bergemann12}, these lines are instead strongly
affected by deviations from local thermodynamical equilibrium (LTE; see below). 
Of these Sr lines, we could use only that at $\lambda 10327$\,{\AA} because that at
$\lambda 10036$\,{\AA} is very weak and blended and located at the edge of one of the CARMENES orders, and the other, at $\lambda 10914$\,{\AA,}
is in one of the spectral gaps of the instrument\footnote{\url{https://carmenes.caha.es/ext/instrument/}}. 
For the 10327\,{\AA} line, we used the oscillator strength given by \citet{Warner68} and the  Van der Waals parameters provided by
the Anstee, Barklem, and O'Mara theory \citep[see][]{Barklem00}. 
This \ion{Sr}{II} $\lambda$10327\,{\AA} line is apparently free of atomic and molecular blends in our stars. 
In fact, in this spectral region the true spectral continuum can be placed more easily.
Finally, the only useful zirconium lines detected in our spectra are the \ion{Zr}{I} lines at $\lambda\lambda$8063.105 and 8070.115\,{\AA}. 
These lines are rather weak but not particularly affected by TiO blends in our stars. 
Oscillator strengths for these lines were taken directly from the VALD3 database. Unfortunately, no yttrium line useful for abundance
analysis was detected in the available spectral range.

\subsection{Spectral synthesis}

We firstly tested our line list by fitting  the observed spectra of the Sun and Arcturus \citep{rei16,Hinkle95} in the
spectral ranges of interest. 
In these stars, the \ion{Rb}{I} $\lambda 7800$\,{\AA} line is heavily blended with a \ion{Si}{I} line, for which we
adopted $\log{gf} = -0.75$ from VALD3. 
Synthetic spectra in LTE were computed with the corresponding model atmosphere using the code
{\tt Turbospectrum v19.1\footnote{\url{https://github.com/bertrandplez/Turbospectrum2019/}}} \citep{ple12}. 
For the Sun, we used a MARCS atmosphere model with parameters $T_{\rm eff}/\log{g}/\xi =5770/4.44/1.1$, and for Arcturus a
model atmosphere with parameters according to \citet{ryd09}. 
Theoretical spectra were convolved with a Gaussian kernel according to the resolving power of the observed spectra and to
mimic the macroturbulence parameter. 
For the Sun, we obtained an LTE Rb abundance of $\log{\epsilon({\rm Rb})} = 2.47\pm0.05$ (average value of the two lines).
This nicely agrees with the value derived by \citet{Grevesse15}. 
For Arcturus, the Rb abundance derived was $2.01\pm 0.01$, which is also in very good agreement with the abundance
obtained by \citet{tom99} and \citet{yon06} of namely $2.02\pm 0.02$. 
A similar test was made for Zr from the selected \ion{Zr}{I} lines. 
Also, nice agreement in the abundance of Zr was found with the values derived by \citet{Grevesse15} and \citet{pet93}
in the Sun and Arcturus, respectively, within $0.05$\,dex. 
In the case of Sr, the LTE abundance derived from the \ion{Sr}{II} $\lambda10327$\,{\AA} is significantly larger
than that usually accepted in the Sun and Arcturus, suggesting deviations from LTE in the formation of this line. 
Eventually, we adopted the solar LTE abundances recommended by \citet{lod19} for Rb (2.47) and Zr (2.58) which nicely agree
with the values obtained here (for Sr, see below). 
We also used the solar photospheric abundances  recommended by this author  for the rest of the elements.

\begin{figure*}
   \centering
   \includegraphics[height=\textwidth,angle=-90]{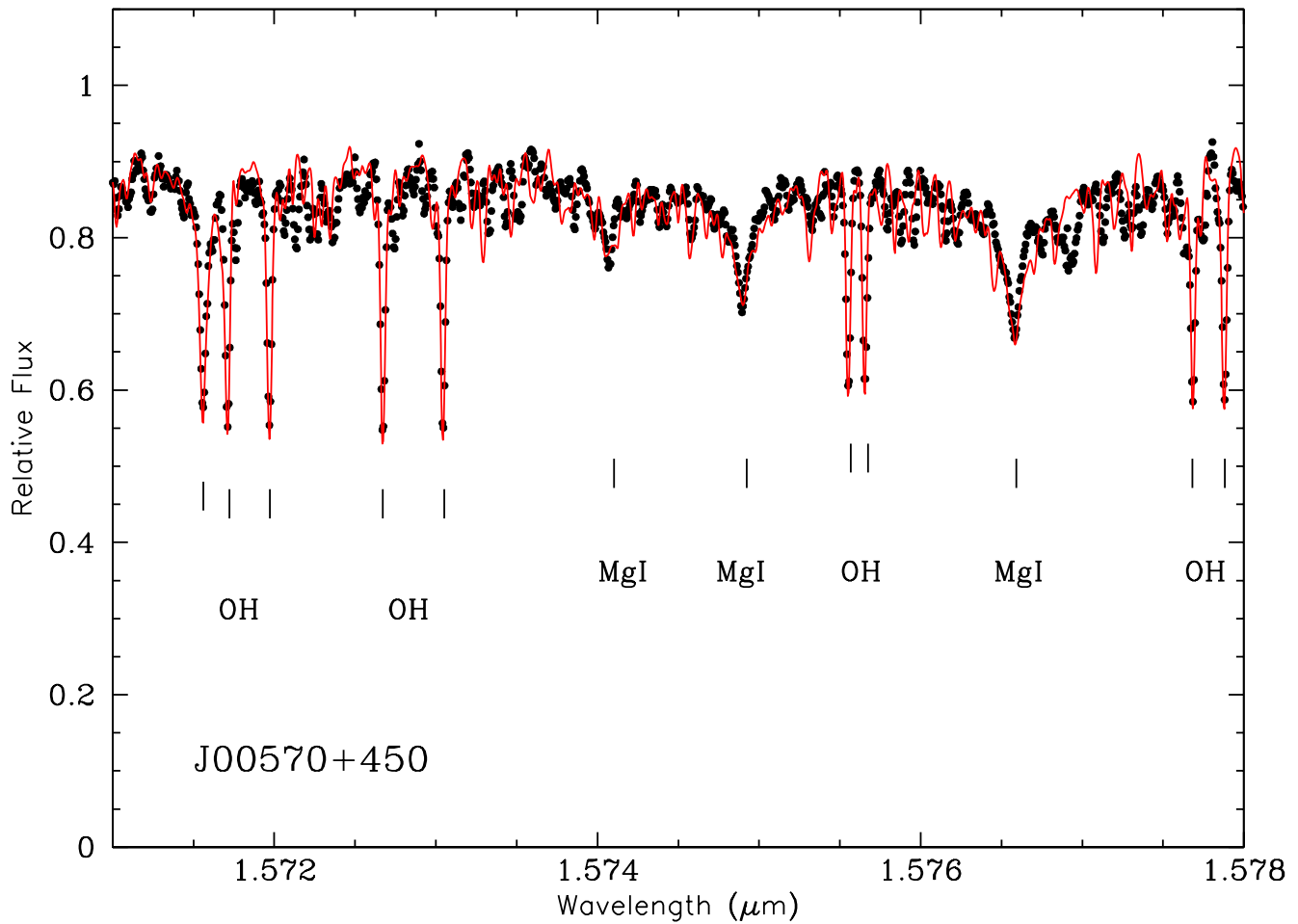}
   \caption{Comparison of the observed (black dots) and synthetic (solid red line) spectra for the M3.0\,V star J00570+450 (\object{G~172--30})
     in the spectral region around $\lambda\lambda 15750\,{\AA}$. 
     Some OH lines, from which the O abundance was derived, and less intense \ion{Mg}{i} lines near $\lambda\lambda\lambda 15740, 15749$,
     and $15765\,\AA$ are marked and labelled. 
   In this region the pseudo-continuum is reduced mainly due to the contribution of an H$_2$O veil.} 
   \label{fig:OH+MgI}
\end{figure*}

The spectral synthesis method for abundance analysis should be applied preferentially when the true continuum can be defined. 
For this reason, it may be difficult to apply to M dwarfs, for which the true continuum cannot be defined in general. 
In these stars, the presence of strong TiO absorption and a {veil} of numerous weak lines of H$_2$O in the VIS and NIR spectral ranges,
respectively, depress the continuum level by some parts per hundred. 
However, in our case an upper envelope of the spectrum can be well defined by connecting the highest flux points in the spectrum
in all the stars. This {pseudo-continuum} can then be relocated with the help of the synthetic spectrum itself. 
To place this pseudo-continuum, the estimation of C/O  is critical, because  this ratio determines the intensity of TiO
and H$_2$O lines for a given set of stellar parameters. 
We followed the iterative method described below.

\begin{itemize}
       \item[(a)] 
         First, a synthetic spectrum was computed for each star with the carbon and oxygen abundances scaled to the metallicity of the model
         atmosphere (see above and \citealt{pass19}). 
         However, for the star J13450+176 (\object{BD+18~2776}; [M/H] $=-0.5$), we considered an oxygen enhancement in the model
         atmosphere, namely, [O/Fe] $=+0.4$, a value typically found in G dwarfs of similar metallicity. 
We always kept the C abundance constant. 
Unfortunately, the CO and CN lines in the covered spectral ranges  are too weak to determine the C abundance. 
Nevertheless, we checked that variations of the C abundance (within $\sim0.25$\,dex) have almost no effect on the synthetic spectrum in the
spectral ranges of interest. 
To compute this initial theoretical spectrum, we did not include the atomic and OH lines in the synthesis. 
In a similar way, theoretical spectra were convolved with a Gaussian function according to the spectral resolution and to mimic the
macroturbulence parameter.
 The stars Karmn 
J11201--104 (\object{LP~733--099}) and J15218+209 (\object{OT~Ser}), which show broader lines, were left out of this convolution. 
For these stars, the observed spectrum was better fitted with a rotational profile. 
Indeed, \citet{Rei18} quoted $v\sin{i} > 2$\,km\,s$^{-1}$ for these stars. 
For them, we used a rotational profile with FWMH in the range 4--6\,km\,s$^{-1}$. 
The profiles of the Rb lines in these stars and in J18174+483 (\object{TYC 3529--1437--1}) may also be affected by the magnetic
field (see below).
    \item[(b)] 
      We then compared the location of the pseudo-continuum in the NIR region with that inferred from the theoretical spectrum itself
      (mainly determined by the H$_2$O veil).
In general, corrections smaller than about 3\,\% were needed. 
We then included the atomic and OH lines in the synthesis and estimated the O abundance by fitting  selected OH lines, in particular those
in the region between 1.57 and 1.59\,$\mu$m (see Fig.~\ref{fig:OH+MgI}). 
The O abundance derived from these selected OH lines should be compatible with the metallicity obtained from the nearby metallic lines
(within its uncertainty of $\pm 0.19$\,dex according to \citet{pass19}; see Fig.~\ref{fig:OH+MgI})\footnote{For unevolved, near-solar
  metallicity stars, [O/Fe] $\approx 0.0$.}.
This agreement  also served as an additional test for the reliability of the pseudo-continuum location.

    \item[(c)] 
      With the O abundance derived in this way, we then compared the observed and calculated intensity of TiO bands in the VIS region, in particular
      the band at $\lambda > 7000$\,{\AA}. 
Adjustments to the O abundance adopted in the synthetic spectrum were made when needed; in any case, they never exceed $\sim 0.20$\,dex. 

    \item[(d)] 
      Finally, with this new O abundance, we fitted again the OH lines in the NIR region and repeated the procedure until we achieved convergence
      (within $\sim 0.1$ dex) between the O abundances estimated from both spectral ranges. Table~\ref{par_ab_data} shows the final C/O  derived
      in our stars. As expected, most of them are very close to the solar ratio (0.575, according to \citet{lod19}), but a few stars show
      significantly lower  C/O or higher than this value.

\end{itemize}

Once the C/O ratio was determined, the next step in the analysis was the determination of the stellar metallicity in a more accurate way.
As mentioned above, the initial metallicitiy was taken from \citet{pass19}. 
However, these metallicities were derived from metallic lines that have moderate (or large) intensity in the atmospheres of M dwarfs and
therefore might be affected by saturation effects. 
A better determination of the metallicity was therefore desirable. 
To do that, we used a number of weak metallic lines available in the spectral range close to the Rb, Sr, and Zr lines. 
The specific lines were
the \ion{Ca}{I} line at $\lambda$10343.819\,{\AA}; 
the \ion{Ti}{I} lines at $\lambda\lambda$ 7791.349, 7949.152, 8068.239, and 8069.799\,{\AA}; 
the \ion{Fe}{I} lines at $\lambda\lambda$7802.473, 7941.088, 7945.846, and 10340.885\,{\AA}; 
and the \ion{Ni}{I} lines at $\lambda\lambda$7788.930 and 7797.580\,{\AA}. 
The weakness of these lines (except that of \ion{Ca}{I}) should minimise saturation effects and possible deviations from LTE. 
In addition, their proximity to the heavy element (Rb, Sr, and Zr) lines may reduce systematic effects introduced by the uncertain location of
the pseudo-continuum when deriving the elemental ratios with respect to the average metallicity ([X/M]), since this uncertainty should cancel out. 
Figures~\ref{fig:ZrI+SrI} and \ref{fig.4} show synthetic fits to some of these metallic lines. 
In most of the stars we found good agreement between the metallicity derived from these metallic species, with a dispersion of about $\pm 0.1$\,dex.
Also, in general our metallicities agree within the uncertainty with those derived by \citet{pass19}, initially adopted here. 
Nevertheless, when a difference larger than $0.15$\,dex was found, we recalculated a model atmosphere with the new metallicity and repeated the
derivation of the metallicity until convergence was reached. 
The largest differences with respect to the metallicities by \citet{pass19} are found in stars with super-solar metalliciy ([M/H] $>0.0$). 
For these stars, we found systematically lower metallicities. 
This difference may be due, as mentioned above, to the fact that in metal-rich stars the lines selected by these latter authors are very strong
and possibly more affected by saturation effects. 
Systematic differences in metallicity are found between \citet{pass19} and similar studies in M dwarfs, although they could also be related to
differences in the stellar parameters (see the discussion in \citealt{pass19} for details). 
Table~\ref{par_ab_data} shows the final average metallicity ([M/H]) derived for our stars. 
The overwhelming majority have metallicity close to solar except for the star J13450+176 (\object{BD$+18$~2776}), which is a mildly metal-poor star. 
This classification agrees with the metallicity determined by other studies for this star \citep[e.g.][]{gai14,man15}. Since for near-solar
metallicity stars, it holds that  [Ni/Fe] $\approx$ [Ti/Fe] $\approx$ [Ca/Fe] $\approx 0.0$,
in the following we refer indistinctly to [M/H] or [Fe/H] as the stellar metallicity. 

\begin{figure*}
   \centering
   \includegraphics[width=\textwidth]{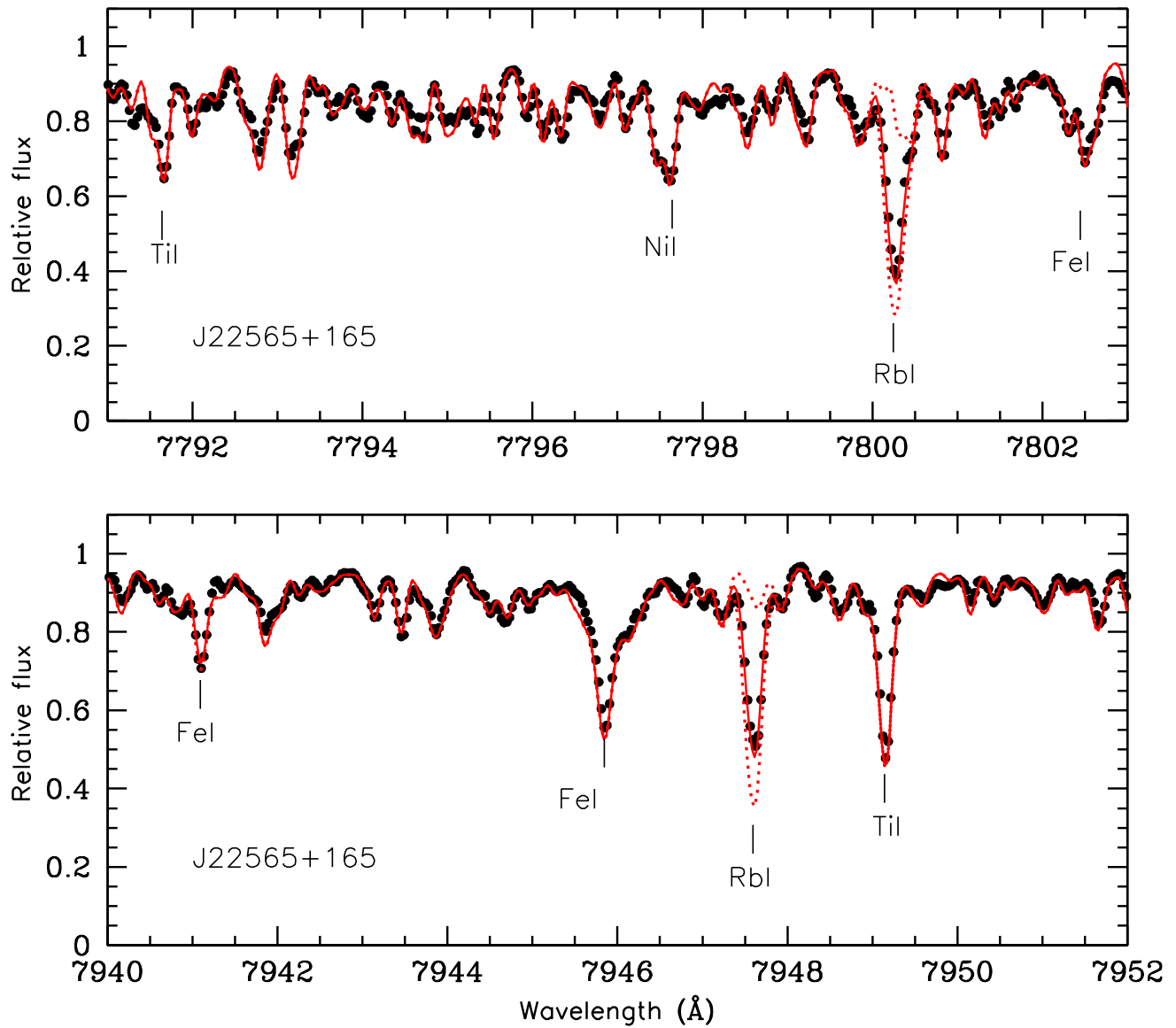}
   \caption{
     Same as Fig.~\ref{fig:OH+MgI} but for the regions around \ion{Rb}{i} $\lambda$7800\,{\AA} ({\em top panel}) and \ion{Rb}{i} $\lambda$7947\,{\AA}
     ({\em bottom panel}) for the M1.5\,V star J22565+165 (\object{HD~216899}). 
   Dash-dotted, solid, and dotted curves show theoretical spectra computed with $\log{\epsilon({\rm Rb})} =$ 2.8, 2.55 (best fit), and no Rb, respectively.
   Some \ion{Ti}{i}, \ion{Ni}{i}, and \ion{Fe}{i} lines are marked and labelled. 
   In these regions, the pseudo-continuum is reduced mainly due to the contribution of TiO.}
  
   \label{fig:RbI}
\end{figure*}

Once C/O  and metallicity were determined, the abundances of Rb, Sr, and Zr were also determined by spectral synthesis fits to the corresponding spectral features. 
Figure~\ref{fig:RbI} shows an example of a theoretical fit (red lines) to the spectral regions of the Rb lines in a representative star of the sample. 
The depression of the continuum mainly due to TiO is particularly apparent in the region of the \ion{Rb}{i} $\lambda7800$\,{\AA} line. 
Fits to some of the metallic lines used for the determination of the average metallicity are also shown.
Similarly, Fig. \ref{fig:ZrI+SrI} shows fits to the \ion{Zr}{I} $\lambda8070$\,{\AA} (left panel) and \ion{Sr}{II} $\lambda$10327\,{\AA} (right panel) lines, respectively. 
The spectral region of the strontium line is apparently not affected by the veil of molecular absorption and therefore the real continuum can be more easily traced. 

\begin{figure*}
    \centering
    \includegraphics[width=\textwidth]{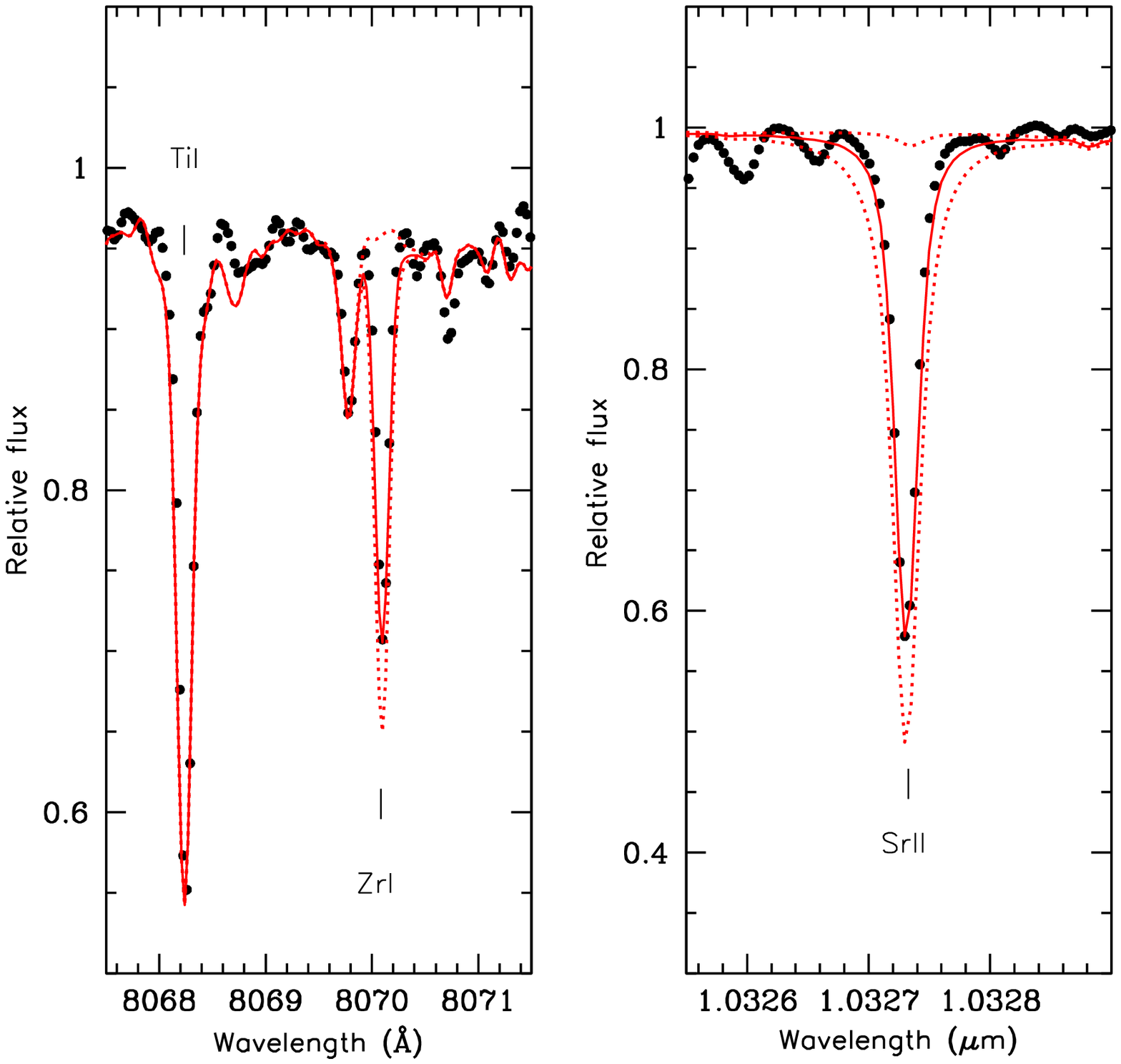}
    \caption{
      Same as Fig.~\ref{fig:OH+MgI} but for the spectral regions of \ion{Zr}{i} $\lambda$8070\,{\AA} ({\em left panel}) and \ion{Sr}{i}
      $\lambda$10327\,{\AA} ({\em right panel}) for the same star.
      Dash-dotted, solid, and dotted curves show theoretical spectra computed with $\log{\epsilon({\rm Zr})} =$ 3.0, 2.75 (best fit), and no Zr,
      respectively, in the {\em left} panel, and with $\log{\epsilon({\rm Sr})} =$ 3.5, 3.2 (best fit in LTE), and no Sr, respectively, in the {\em right} panel.
    }
    \label{fig:ZrI+SrI}
\end{figure*}  

The two main sources of error in the abundances are measurement and analysis error caused by errors in
the adopted model atmosphere parameters. 
The scatter of the abundances provided by individual lines of the same species is a good guide to measurement error.
We found excellent agreement between the Rb abundances derived from both lines, with a dispersion of less than $0.05$\,dex
(see Table~\ref{par_ab_data}). 
This agreement contrasts with the differences ($\sim0.1$\,dex) found by \citet{tom99} and \citet{yon06}, which led them to
exclude the \ion{Rb}{i} $\lambda7947$\,{\AA} line from their analyses, which appeared to be more highly blended. For zirconium,
we also found excellent agreement between the abundances derived from both lines (see again Table~\ref{par_ab_data}).
The error caused by uncertainties in the adopted stellar parameters can be estimated by modifying the stellar parameters by the
quoted errors in the analysis of a typical star in the sample and checking the effect on the abundance derived for each species,
namely: $\pm 60$\,K in $T_{\rm eff}$, $\pm 0.06$\,dex in $\log{g}$, $\pm 0.5$\,km\,s$^{-1}$ in  $\xi$, $\pm 5\,\%$ in C/O, and $\pm 0.15$\,dex in [Fe/H]. 
For a typical M dwarf with parameters $T_{\rm eff}/\log{g}/[{\rm Fe/H}] = 3750/4.7/0.0$, we found that the abundances derived
are mostly affected by the uncertainty in  $T_{\rm eff}$:
$\pm 0.10$, $\pm 0.10$, and $\pm 0.08$\,dex for Rb, Sr, and Zr, respectively. The uncertainties in gravity and microturbulence
have a similar, although minimal impact on the abundances derived for the three elements ($\leq 0.03$\,dex). 
The uncertainty in the metallicity and C/O has a slightly larger effect and is  typically $0.05$\,dex for the three
species. 
Adding these uncertainties together  quadratically with the dispersion around the mean abundance value and the continuum
uncertainty (about 1--2\,\% in the case of Rb and Zr), we estimated a total uncertainty in [X/H] of $\pm 0.12$\,dex for Rb
and Sr, and $\pm0.10$\,dex for Zr. 
For these heavy elements, the abundance of the element relative to average metallicity, [X/Fe], holds the most interest. 
This ratio is more or less sensitive to the uncertainties in the atmospheric parameters depending on whether changes in
the stellar parameters affect the heavy element abundance and metallicity in the same or opposite sense. 
In our case, we estimated total uncertainties of $\pm0.12$, $\pm0.14$, and $\pm0.13$\,dex for [Rb/Fe],
 [Sr/Fe], and [Zr/Fe], respectively.

\subsection{Non-LTE effects in \ion{Rb}{I} and \ion{Sr}{II} lines}

It is well known that the resonant lines of other alkali elements, such as Na or K, are strongly influenced by deviations from LTE \citep{Bruls92}. 
The structure of the energy levels of \ion{Rb}{i} is very similar to those of \ion{K}{i} and \ion{Na}{i} and, by analogy, some non-LTE effects
in the formation of Rb lines are expected. 
As far as we know, no previous study exists on non-LTE effects in \ion{Rb}{I} in the literature. 
Below we briefly describe the method followed to evaluate these effects in M dwarfs. 
A detailed discussion in a wider range of stellar parameters will be presented elsewhere \citep{Korotin20}.

Our rubidium atom model includes 29 levels of \ion{Rb}{i} and the ground level of \ion{Rb}{ii}.
Radiative transitions between higher levels are very weak and should not affect the accuracy of the non-LTE calculation. 
Fine structure was considered only for the 5p~2P$^{0}$ level, since it is closely connected to the most important transitions including the resonance lines. 
In the atmospheres of cool dwarfs, Rb exists almost exclusively in the form of \ion{Rb}{ii}, since its first ionisation potential is  only 4.18\,eV. 
In turn, all excited \ion{Rb}{ii} levels are separated from the main one by more than 16.5\,eV. 
Nevertheless, 180 supplementary levels of \ion{Rb}{i} and 15 levels of \ion{Rb}{ii} were included in the model to assure the conservation of the particle number. 
In total, 160 radiative transitions were taken into account for the calculation of the population levels.
We included collisional transition rates between the first six levels of \ion{Rb}{i} according to \citet{Vainshtein79}. 
For other collisional transitions between levels with energies below 3.9\,eV, calculations from \citet{Park71} were used. 
For the rest of the bound--bound collisional transitions, we adopted the formulae from \citet{Regemorter62} for
permitted transitions and from \citet{Allen73} for the forbidden ones. 
Collisional ionisation rates were calculated using the formula from \citet{Seaton62}. 
Collisions with hydrogen for the first nine levels were described using the detailed quantum mechanical calculations from \citet{Yakovleva18}. 
For other levels, the classical Drawin formula was used in the form suggested by \citet{Steenbock84} with the correction factor $S_{H} = 0.05$.
The population of the atomic levels for \ion{Rb}{i} was determined using the code {\tt MULTI} \citep{Carlsson86} with some modifications following \citet{Korotin99}.

\begin{figure}
    \centering
    \includegraphics[width=11cm]{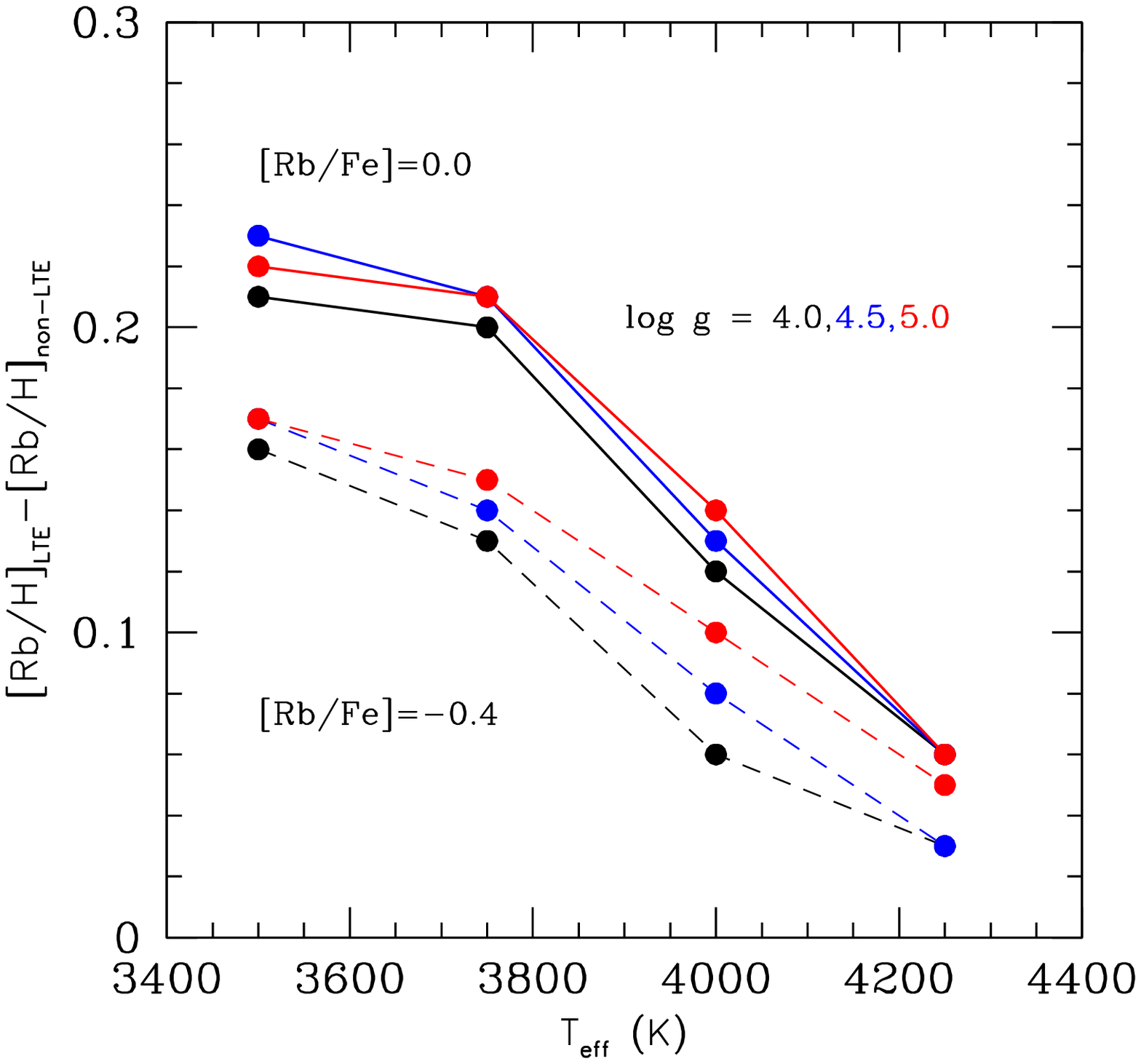}
    \caption{Difference between the LTE and non-LTE Rb abundances derived from the \ion{Rb}{i} $\lambda$7800\,{\AA} line for typical
      $T_{\rm eff}$ and $\log{g}$ (colour coded) in our sample stars.
    Solid and dashed lines correspond to [Rb/Fe] ratios of $0.0$ and $-0.4$, respectively 
    (see text for details).}
    \label{fig.4}
\end{figure}

The non-LTE corrections lead to a strengthening of the resonance lines in cool dwarfs ($\log{g} > 4.0$). 
This strengthening is due to the overpopulation of the ground level because of the so-called {super-recombination} processes. 
These processes are common in other alkali metals and were described in detail in \citet{Bruls92}. 
We first tested our non-LTE calculations for the Rb lines with the Sun. 
The correction was $-0.12$\,dex, and we obtained a solar non-LTE abundance of Rb of $2.35\pm0.05$. It is noteworthy that this
value is, for the first time,  in very good agreement with the recommended meteoritic abundance ($2.36\pm0.03$, \citealt{lod19}). 
We then calculated a grid of non-LTE corrections for the stellar parameters typical of our stellar sample, namely $T_{\rm eff}$ from
3500 to 4500\,K, $\log{g}$ from 4.0 to 5.0\,dex, [Fe/H] from $-0.5$ to $+0.5$\,dex, and [Rb/H] from $-0.4$ to $+0.4$\,dex (see Table~\ref{par_ab_data}). 
In general, the non-LTE correction decreases with increasing temperature since the hydrogen collisions become more important.  
The value of the correction is also determined  by changes in the distribution of the average intensity with wavelength and depth in
the stellar atmosphere, also affecting  the efficacy of the super-recombination processes. 
However, the fraction of \ion{Rb}{I} below $\sim 4500$\,K  increases and the intensity of the resonance lines increases sharply. The
resonance lines are formed mainly in the highest layers of the photosphere where the density is the lowest and the collision rates
cannot compensate the imbalance introduced by radiative processes, and therefore the non-LTE correction increases. 
This effect is enhanced with increasing [Rb/Fe] ratio (see Fig.~\ref{fig.4}).  
Non-LTE abundance corrections range from $-0.25$ to $-0.03$\,dex, being $\sim -0.15$\,dex on average for the typical atmosphere
parameters of our stars.
Figure~\ref{fig.4} shows an example of the non-LTE abundance corrections to the \ion{Rb}{I} $\lambda$7800\,{\AA} line for specific
atmosphere parameters and derived [Rb/Fe] ratios. 
As expected, the non-LTE correction decreases with increasing $T_{\rm eff}$ and decreasing $\log{g}$. 
However, for the \ion{Rb}{I} $\lambda$7947\,{\AA} line,  corrections are slightly lower for a given set of atmosphere parameters and Rb
abundance because this line forms slightly deeper in the atmosphere. 
Again, a more detailed discussion will be presented in \citet{Korotin20} including  non-LTE corrections in giant stars.
In passing, we note that our non-LTE correction for Rb in Arcturus is almost zero.

To evaluate the non-LTE effects in the formation of the \ion{Sr}{II} line in the NIR, we used the atom model proposed by \citet{Andrievsky11}. 
We refer to this author for a more detailed description of the atomic levels used and non-LTE calculations. 
Briefly, collisional transitions caused by inelastic collisions with \ion{H}{I} were considered according to the formulae of
\citet{Drawin68}  as compiled by \citet{Steenbock84} with a correction factor of $S_H=0.01$.
This correction factor was obtained by fitting the line profile of different multiplets in the solar spectrum, and was confirmed
by the analysis of \citet{Bergemann12}.
Atomic level populations were also determined using the code {\tt MULTI} with modifications according to \citet{Korotin99}. 
The atomic model used here describes correctly the profiles of the Sr lines in the blue and NIR spectrum of the Sun, giving
an identical Sr abundance (2.92; this is the value adopted in this study as the solar Sr abundance). 
For Arcturus, the non-LTE analysis   gave a very similar Sr abundance, namely $2.23 \pm 0.04$\,dex from the three NIR lines, which
is significantly lower than the LTE value of 2.40 from \citet{pet93}. 
It should be noted that non-LTE effects in Sr lines differ significantly from line to line. 
For instance, in the spectrum of the Sun the resonance lines ($\lambda\lambda 4077.71$ and 4215.52\,\AA) form almost in LTE, the
two subordinate blue lines ($\lambda\lambda 4161.79$ and 4305.44\,\AA) show weak non-LTE corrections ($\sim +0.05$\,dex), while
for the NIR lines non-LTE corrections are severe (from $-0.20$ to $-0.36$\,dex). 
Unfortunately, of these three lines, only that at $\lambda10327$\,\AA\ is covered by CARMENES. 
Similarly to Rb, we calculated a grid of non-LTE abundance corrections within the range of the atmospheric parameters of our
stars and Sr abundances in the range $-0.3\leq$ [Sr/H] $\leq +0.3$. 
Non-LTE abundance corrections are also negative and range from $-0.28$ to $-0.13$ dex, with the correction decreasing with increasing
temperature and gravity. 
The correction is larger for moderately metal-poor stars due to the smaller contribution of collisions in the statistical equilibrium. 
The non-LTE abundances of Sr derived are shown in Table~\ref{par_ab_data} (we prefer not to include the non-LTE Rb abundances
in this table for succinctness. The effect is abundances typically lower by  $\sim 0.15$ dex.

\section{Results and Discussion}
\label{discussion}

\begin{figure}
    \centering
    \includegraphics[width= 9.5 cm]{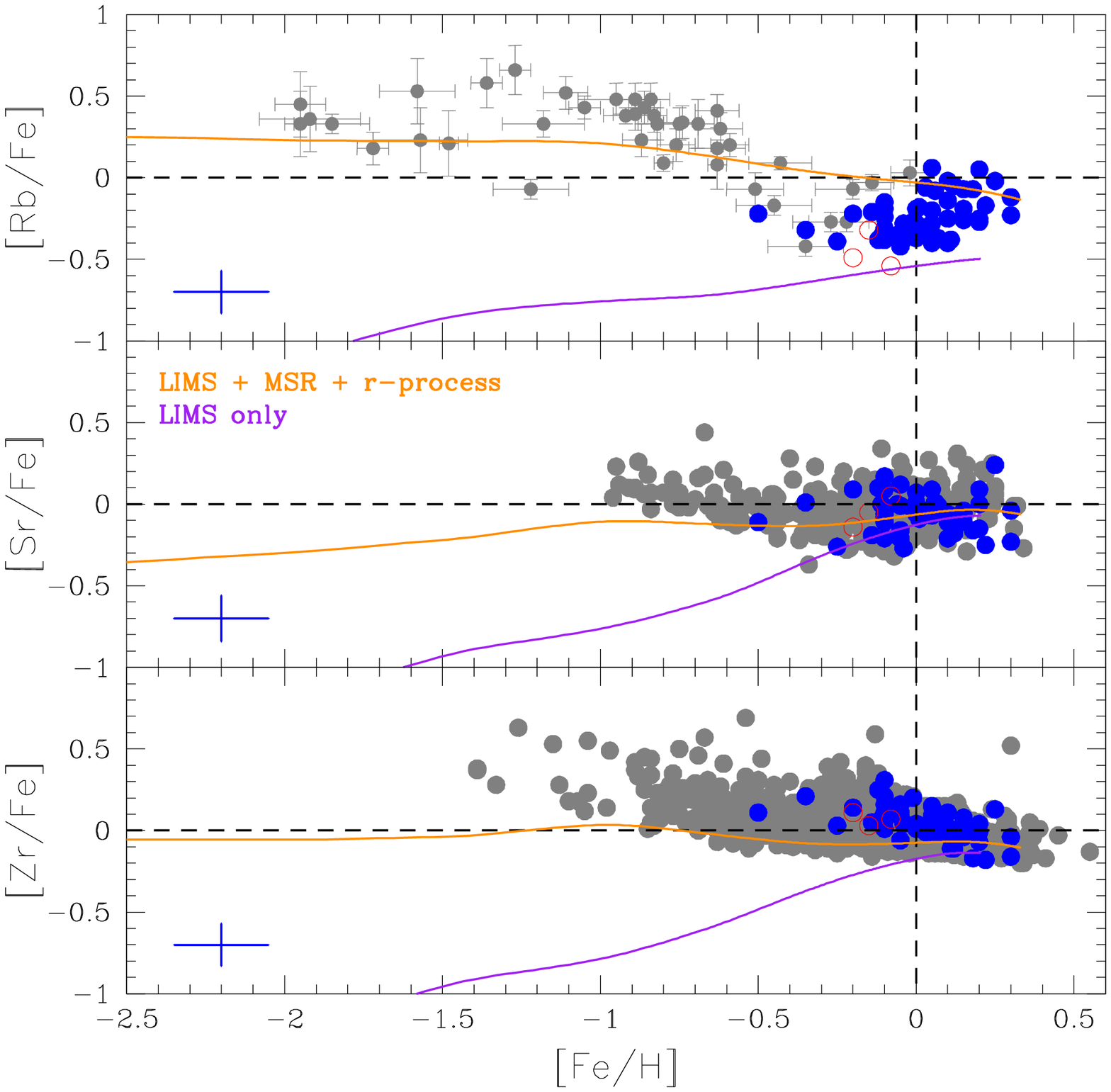}
    \caption{[Rb,Sr,Zr/Fe] vs. [Fe/H] diagrams, from top to bottom.
      In the three panels, blue filled circles are our program stars. In the top panel, grey dots with error bars are the [Rb/Fe] ratios
      derived in halo and disc giant and dwarf stars by \citet{gra94} and \citet{tom99}. In the middle and bottom panels, grey dots are
      the [Sr/Fe] ratios derived by \cite{mis19}, and [Zr/Fe] ratios by \citet{bat16} and \citet{del17}, both in thin disc dwarf stars. A t
      ypical error bar in the [X/Fe] ratios for our M dwarfs is shown in the bottom left corner of each panel. For dwarf and giant stars
      (grey dots) the error bars are slightly smaller, which we have omitted for clarity.  The red open circles in all the panels are the
      stars in which the Rb lines may be affected by magnetic field. Continuous lines are theoretical GCE predictions by \citet{pra18,pra20}. Orange
      lines include the contributions from low- and intermediate-mass stars (LIMS), rotating massive stars (MSR), and the r-proces. Magenta
      lines include only LIMS. }
    \label{fig:[X/Fe]}
\end{figure}

Table~\ref{par_ab_data} shows the Rb, Sr, and Zr abundances determined in our sample of stars.  As is customary, the abundances
in Table~\ref{par_ab_data} are plotted as [X/Fe] against [Fe/H] in Fig.~\ref{fig:[X/Fe]} to reveal trends in the relative
abundance of element X and the metallicity. 
Two points are immediately apparent: 

\begin{itemize}
    
    \item[(1)] 
      The [Sr,Zr/Fe] ratios derived in M dwarfs (blue dots) nicely agree with those derived in FGK dwarfs (grey dots) for the metallicity
      range studied here. 
      Data in FGK dwarf stars were taken from \citet{bat16}, \citet{del17}, and \citet{mis19}, which we considered as representative studies
      in the recent literature based on very high S/N spectra (formal errors in [Fe/H] and [Sr,Zr/Fe] in these studies were typically
      $0.06$\,dex and $\sim 0.10$\,dex, respectively). 
For the range of metallicity studied, the [Sr/Fe] ratio  clusters around the solar value, the average ratio being $-0.06\pm 0.10$\,dex. 
The [Zr/Fe] ratio shows a small trend of increasing ratio for decreasing metallicity, identical to that found in FGK dwarfs (grey
dots), although observations indicated that this trend flattens in metal-poor stars ([Fe/H] $< -1.5$ (see e.g.
\citealt{bre06,red06})\footnote{It is beyond the scope of this study to discuss the abundance trends in metal-poor stars.}. 
For [Zr/Fe] the average ratio is $0.06\pm 0.09$\,dex. 
In both cases (Sr and Zr), the dispersion around the mean is lower than the expected observational errors in the [X/Fe] ratios. 
 
    \item[(2)] 
The [Rb/Fe] ratios derived in M dwarfs are systematically below the solar value. 
This deficiency is on average [Rb/Fe] $=-0.26 \pm0.13$\,dex in the metallicity range studied, and would remain if non-LTE Rb abundances
were used instead because the systematic correction by $\sim 0.15$\,dex to lower abundances would be almost compensated by a lower solar
Rb abundance (2.35; see above) when deriving the [Rb/Fe] ratio.
Interestingly, Fig.~\ref{fig:[X/Fe]} (top panel) shows that the few [Rb/Fe] ratios derived in FGK dwarfs in the similar metallicity
range also show negative values. 
This behaviour is in contrast with that of its associated elements Sr and Zr. However, the average [Rb/Fe] ratio shows a significant
dispersion, slightly larger than the expected observational errors. 
Furthermore, a hint of an increasing trend in the [Rb/Fe] ratio with metallicity can be seen for [Fe/H] $\gtrsim -0.3$, although this
trend is weak (correlation coefficient of $r=0.43$).

\end{itemize}

The Rb underabundances found in our M dwarfs are unexpected and require further discussion. 
We cannot exclude systematic errors in the analysis, such as the existence of unknown blends in the \ion{Rb}{I} lines,
although there is no evidence of this.
In particular, in our program stars, the strengthening of the line with decreasing $T_{\rm eff}$ is consistent with the
behaviour of a resonance line of a heavy-element neutral species. 
The very well known blend at $\lambda\sim 7800$\,{\AA} with a \ion{Si}{I} line is not expected to play any role here
because of its high excitation energy ($\chi=6.18$\,eV) and the low $T_{\rm eff}$ of our stars. 
Another possibility is incorrect placement of the spectral continuum. 
We explored this possibility by changing its position up (down) by a given fraction. 
However, this led systematically to unrealistically large (small) Ti or/and O abundances from fits to the TiO lines in this region. 
On the other hand, underestimation of  $T_{\rm eff}$ would lead systematically to the derived low Rb abundances, but differences
of at least as large as $\sim 150$\,K would be required to explain this Rb underabundance. 
Although such a systematic error in $T_{\rm eff}$ cannot be fully excluded, this would also imply a significant change in the
derived Sr and Zr abundances, which would be translated into a disagreement between the [Sr,Zr/Fe] ratios in M and FGK dwarfs (see Fig.~\ref{fig:[X/Fe]}). 
Systematic errors in the other stellar parameters at the level of the expected uncertainties have only a minor impact on the derived Rb abundance. 
In this respect, the most important parameter is C/O. 
However, we verified that unrealistically low O and C abundances would be required  to remove the observed Rb deficiency (which
would also affect the fits of the TiO lines in the Rb spectral region). 

We  also explored possible correlations of the Rb abundance with the adopted $T_{\rm eff}$ and $\log{g}$. While there is no correlation
with $T_{\rm eff}$, we found a weak correlation with $\log{g}$, namely larger Rb abundances for decreasing $\log{g}$ (correlation coefficient $r=0.44$). 
This correlation, although weak, could be related with the stellar age. 
Evolutionary tracks show that M dwarfs of $Z\sim 0.0$ and $\log{g} \lesssim 4.6$\,dex may have young ages because they are still
contracting quasi-hydrostatically towards the main sequence \citep{bre12,tan14,bar15}.
Effects of low gravity in spectra of very young M (and L) dwarfs have been observed in stellar kinematic groups and
the 125\,Ma-old Pleiades open cluster \citep{moh03, cru09, shk09, zap14, Alo15}.
Interestingly, the overwhelming majority of the stars in our sample with $\log{g} < 4.7$\,dex have [Fe/H] $>+0.15$\,dex and
therefore may be relatively young. However, stellar chromospheric activity is observationally known to decline steeply with stellar
age for solar-type field dwarfs that are younger than the Sun, and thereafter continues to decrease very slowly as stars get significantly older \citep{sod10,ram14}. 
This seems to  also be the case for M dwarfs \citep[see e.g.][]{delf98,Piz03,as17} 
and, since the resonance Rb lines form in the upper layers of the atmosphere, the formation of these lines could be affected by
the stellar activity and thus lead to incorrect estimation of the Rb abundance. 

\begin{figure}
    \centering
    \includegraphics[width= 9. cm]{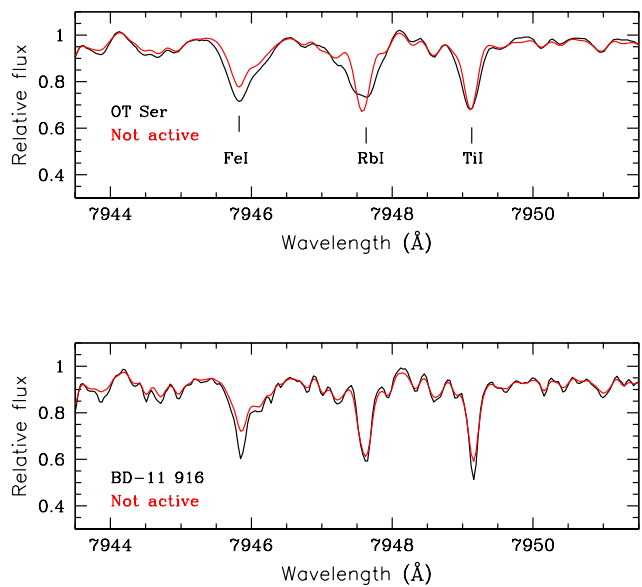}
    \caption{Normalised spectra around the \ion{Rb}{I} $\lambda7947$\,{\AA} line of stars with strong (J15218+209, OT Ser, M1.5\,V,
      {\em top}, black line) and faint magnetic field (J04376--110, \object{BD-11~916}, M1.5\,V, {\em bottom}, black line), in comparison
      with the spectra of a CARMENES GTO, slowly rotating inactive star of the same spectral type (J16254+543, \object{GJ~625}, M1.5\,V, red line)
      broadened to the corresponding target star $v \sin{i}$.}
    \label{fig:active}
\end{figure}

Related to the stellar ages above, another possible effect to take into account for very active stars  with strong magnetic fields
is the Zeeman broadening of the spectral lines with large Land\'e factors. The affected lines are shallower when compared to those
of a non-active star, and this can affect the abundance determination by spectral synthesis if the magnetic field is not included. 
We identified three stars in our sample, namely J11201--104 (\object{LP~733--099}), J15218+209 (\object{OT~Ser}), and
J18174+483 (\object{TYC 3529--1437--1}), with strong activity levels based on their H$\alpha$ emission from the CARMENES
activity indicator analysis \citep{scho19}. By comparison with non-active stars of the same spectral type, we confirmed
that in the three cases both  Rb lines are indeed affected by Zeeman broadening (Fig.~\ref{fig:active}). 
For these three stars, the Rb abundances are amongst the lowest derived in the sample, and are consequentially highly
uncertain (see Table~\ref{par_ab_data}).We plot them with red open circles in Figs.~\ref{fig:[X/Fe]} and \ref{fig:[Rb/X]}. 
Nevertheless,  even excluding these active stars and those with $\log{g}\lesssim 4.7$\,dex from our analysis, the [Rb/Fe] deficiency
and the positive correlation with [Fe/H] still remains (showing a slightly higher correlation, $r\sim 0.47$). On the other hand, no
correlation is found between the derived Sr and Zr abundances with $T_{\rm eff}$ and $\log{g}$. 
 
There could be other possibilities. Indeed, 
 \citet{lam95} found a similar problem when deriving Rb abundances in evolved cool M giants. 
 These latter authors obtained systematic Ti underabundances from TiO lines in the region of the Rb lines, which is at
 odds with the metallicity derived from atomic lines. They discussed different possibilities to explain this effect, but
 could only solve this problem by adding an extra quasi-continuous opacity $\kappa$(TiO) in the spectral synthesis. 
 They also showed that the Rb abundance expressed as [Rb/Fe] was almost insensitive to the chosen combination of quasi-continuous
 opacity and Ti abundances as long as [Fe/H] is derived from metallic lines around the \ion{Rb}{I} lines (see Section~\ref{sample}). 
 However, we did not find this issue here, as the metallicty inferred from the TiO lines around the Rb spectral region is compatible
 (within the uncertainty) with that derived from individual metallic lines in the other spectral ranges.
Therefore, other possibilities have to be explored. 
For instance, it is well known that the Rb abundance in the Sun is rather controversial \citep{lod19}. 
Initial determinations by \citet{wit71}, \citet{ros76}, \citet{and89}, and \citet{pal03}, among others, obtained 2.60 with very
little dispersion. 
This value is significantly higher than the meteoritic value derived from CI chondrites ($2.36\pm 0.03$; e.g. \citealt{lod09}). 
Later, \citet{Grevesse15} reduced this discrepancy and determined $ 2.47\pm 0.07$\,dex using a 3D atmosphere model for the Sun,
but still $\sim 0.1$\,dex larger than the meteoritic value.  
While our non-LTE abundance ($2.35\pm 0.02$\,dex; see Section~\ref{sample}) is in nice agreement with the meteoritic value, it
seems that the issue may no longer be with the Rb in the Sun, but perhaps with the Sun in respect to the nearby stars. 

There is some observational evidence that Rb in the Solar System could be overabundant with respect to the local ISM. 
Neutral Rb is expected to follow the spatial trend of neutral K (another alkali element) in interstellar clouds because these
latter have similar ionisation potentials and chemical properties and therefore a determination of the interstellar elemental
Rb/K ratio may provide insight into this issue. 
While the alkali ratios Li/K and Na/K measured in the local ISM \citep{wel01} are consistent with those in the Solar System,
this is not the case for Rb/K. 
\citet{wal09} studied Rb/K toward seven different ISM directions from the Sun and obtained values $\sim 34$\,\% lower than
the meteoritic (solar) value, except perhaps towards $\rho$~Oph~A. 
The condensation temperature for Rb is lower than that for K.
As a result, if  depletion onto grains was the only deciding factor then one would expect a higher Rb/K, contrary to the observational results. 
Depletion patterns in the local ISM are not always found for elements with similar condensation temperatures to that of Rb.
For example, \citet{car06} found a depletion by $\sim 25$\,\% in Ge with respect to the meteoritic value but no evidence for Cd and Sn. 
This result led \citet{wal09} to suggest that neutron capture elements synthesised by the weak s-process and the r-process in massive
stars (as Rb) appear to show significantly lower interstellar abundances  than seen in the Solar System when compared with those
synthesised mainly in low-mass stars. 
However, \citet{rit18}, in a more comprehensive study on the depletion pattern of several neutron-capture elements
(Ge, Ga, As, Kr, Cd, etc.) in the local ISM, concluded that a simple dichotomy between the production of heavy elements by
low- and high-mass stars cannot account for the unexpectedly low interstellar abundance of Rb and other neutron capture elements.
The value found here for M dwarfs, if confirmed, may open this discussion again. 
To elucidate this, K abundance determinations are suggested in M dwarfs.
Unfortunately, most \ion{K}{I} lines identified in the CARMENES spectra of M dwarfs are very strong and are affected by magnetic
activity (Fuhrmeister et~al., in~prep.). Accepting the [Rb/Fe] discussed above as genuine, in the following section we compare our
observational results with a Galactic chemical evolution (GCE) model for neutron-capture elements in order to extract information
about possible  nucleosynthetic implications.

\subsection{Comparison with GCE models}

Figure 5 compares the observed [Rb,Sr,Zr/Fe]-versus-[Fe/H] trends with the theoretical predictions according to the GCE
model from \citet{pra18,pra20} for the solar neighbourhood. 
This model is a state-of-the-art one-zone GCE model as far as the neutron capture elements are concerned, which uses recent
yields from LIMS \citep{cri15}, as well as of massive rotating stars \citep[MSRs;][]{lim18}.
The former are the main contributors to neutron-capture elements formed through the main component of the s-process, while
the latter contribute through the weak s-process, for which the inclusion of stellar rotation becomes critical. The r-process
is assumed in this GCE model to be associated to the evolution of massive stars. 
For Rb, at the time of the Solar System formation, this GCE model predicts s- and r-fractions of $51\,\%$ and $49\,\%$ ($33\,\%,
67\,\%$ for $^{85}$Rb; $96\,\%, 4\,\%$ for $^{87}$Rb), and $91.2\,\%, 73\,\%, 82\,\%$ (s-fraction) and $8.3\,\%, 22\,\%$, $18\,\%$
(r-fraction) for the light heavy elements Sr, Y, and Zr, respectively\footnote{There is a 0.5 \% fraction contribution for Sr due to the p-process.}. 
We note that the model accounts for the observed Solar System abundances of most of the heavy elements (Z$>30$) within $10\,\%$. 
A detailed description of this GCE model, as well as of the stellar yields used, can be found in \citet{pra18,pra20}. 

We compare the model predictions with observations  in two illustrative cases in Fig. 5: 
the orange line shows the predicted relationships when all the nucleosynthetic sources are included (LIMS, MSR, and r-process), which
we refer to as the `baseline' model, while the magenta line shows the result when only the LIMS contribution is considered
(i.e. the main s-process contribution). 
The baseline model considerably    better reproduces the average observed [Rb/Fe]-versus-[Fe/H] trend at low metallicity
([Fe/H]$\lesssim -1.0$; top panel of Fig. 5),
where the [Rb/Fe] ratio is almost flat ($\sim+0.3$\,dex; i.e. it has a primary behaviour), similar to that observed for a genuine
r-process element (e.g. Eu; \citealt{mas03,zha16,del17}). 
We conclude that at low metallicity the r-process is the main source of Rb, although with a non-negligible contribution from
the weak s-process occurring in MRS. The Rb production from LIMS has a secondary behaviour (magenta line) and only becomes significant
at [Fe/H] $\gtrsim -0.3$ (see Fig. 5, top panel).
At [Fe/H] $\gtrsim -1.0$, the observed [Rb/Fe] ratio declines in a similar way as other primary elements (or elements with a
main r-process origin) due to the increasing contribution to Fe from type Ia supernovae. 
However, contrary to what is typically observed for these elements, the [Rb/Fe] ratio reaches $\sim 0.0$ at a lower metallicity
and is below the solar ratio at [Fe/H]$\approx 0.0$ according to our observational results. 
We note that \citet{gra94} and \citet{tom99} also derived [Rb/Fe] $<0.0$ in a few stars with [Fe/H] $>-1.0$\footnote{We have
  scaled the [Rb/Fe] ratios derived by these authors to the solar Rb abundance (2.47) adopted here.}. 
This is an unexpected result and has not been observed in any other heavy element. Typically most of the heavy elements show
average solar ratios ([X/Fe]) at [Fe/H] $\sim 0$, but not systematically below solar, as can be seen, for instance, in the [Sr,Zr/Fe]-versus-[Fe/H]
relationships (middle and lower panels of Fig.~\ref{fig:[X/Fe]}). 
The observed behaviour of the [Sr,Zr/Fe] ratios is reasonably well reproduced by our baseline model. 
However, this baseline model cannot reproduce the observed [Rb/Fe] ratios at [Fe/H] $\gtrsim -1.0$. 
This fact would imply at least two things: 
($a$) the Rb production in massive stars is overestimated in the GCE model at these metallicities; 
and ($b$) because of the time delay of the AGB stars producing cosmic Rb, their contribution is theoretically predicted to
start contributing at metallicities higher than [Fe H]$\sim -0.7$ \citep{bis14,pra18}. 
However, as seen in Fig.~\ref{fig:[X/Fe]}, the contribution from AGB stars  
is not enough to explain the observed [Rb/Fe] ratios at nearly solar metallicity.
We note that Rb yields  from different AGB stellar models \citep[e.g.][]{kar14} may be systematically larger with respect to those
by \citet{cri15} adopted in \citet{pra18}. These differences are  due to different prescriptions used for mixing and mass-loss
rates. However, in this case the yields of other heavy elements produced by the s-process, such as Sr, Y and Zr in particular,
are  also enhanced, resulting in a probable overproduction of these elements, which is not required according to our GCE model
(see Fig. 5, middle and bottom panels).
\begin{figure}
   \centering
   \includegraphics[width=9 cm]{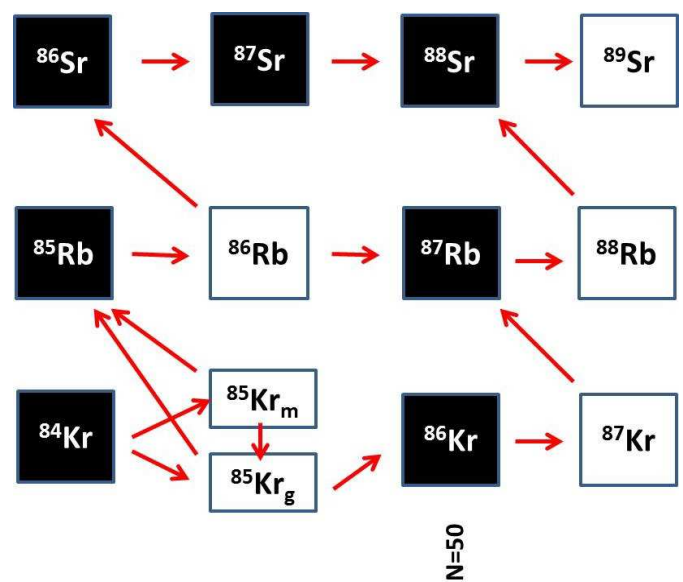}
   \caption{Scheme illustrating the operation of the two branches at $^{85}$Kr (ground state) and $^{86}$Rb that
     determine the Rb production by the s-process. Black filled nuclei are stable, while the empty ones are unstable.
     Arrows show possible nuclear reactions, particularly neutron captures or $\beta$-decays. In case of low neutron
     density ($N_n<10^9$ cm$^{-3}$), both branches are closed and the s-process follows  the chain $^{84}\rm{Kr} \to ^{85}\rm{Kr}
     \to ^{85}\rm{Rb} \to ^{86}\rm{Rb} \to ^{86}\rm{Sr}$. Owing to the large neutron-capture cross section of $^{85}$Kr compared
     to those of the neutron-magic nuclei (N=50) $^{88}$Sr, $^{89}$Y,
     and $^{90}$Zr, a low amount of Rb is produced. On the contrary, at higher neutron density, the two branches
     are open and the s-process sequence deviates toward $^{86}$Kr and $^{87}$Rb, both of which are neutron magic
     nuclei, and therefore both present very low
     neutron-capture cross sections. In the latter case, large overabundances of Kr and Rb are found. } 
    \label{fig:branches}
\end{figure}

\begin{figure}
   \centering
   \includegraphics[width=9. cm]{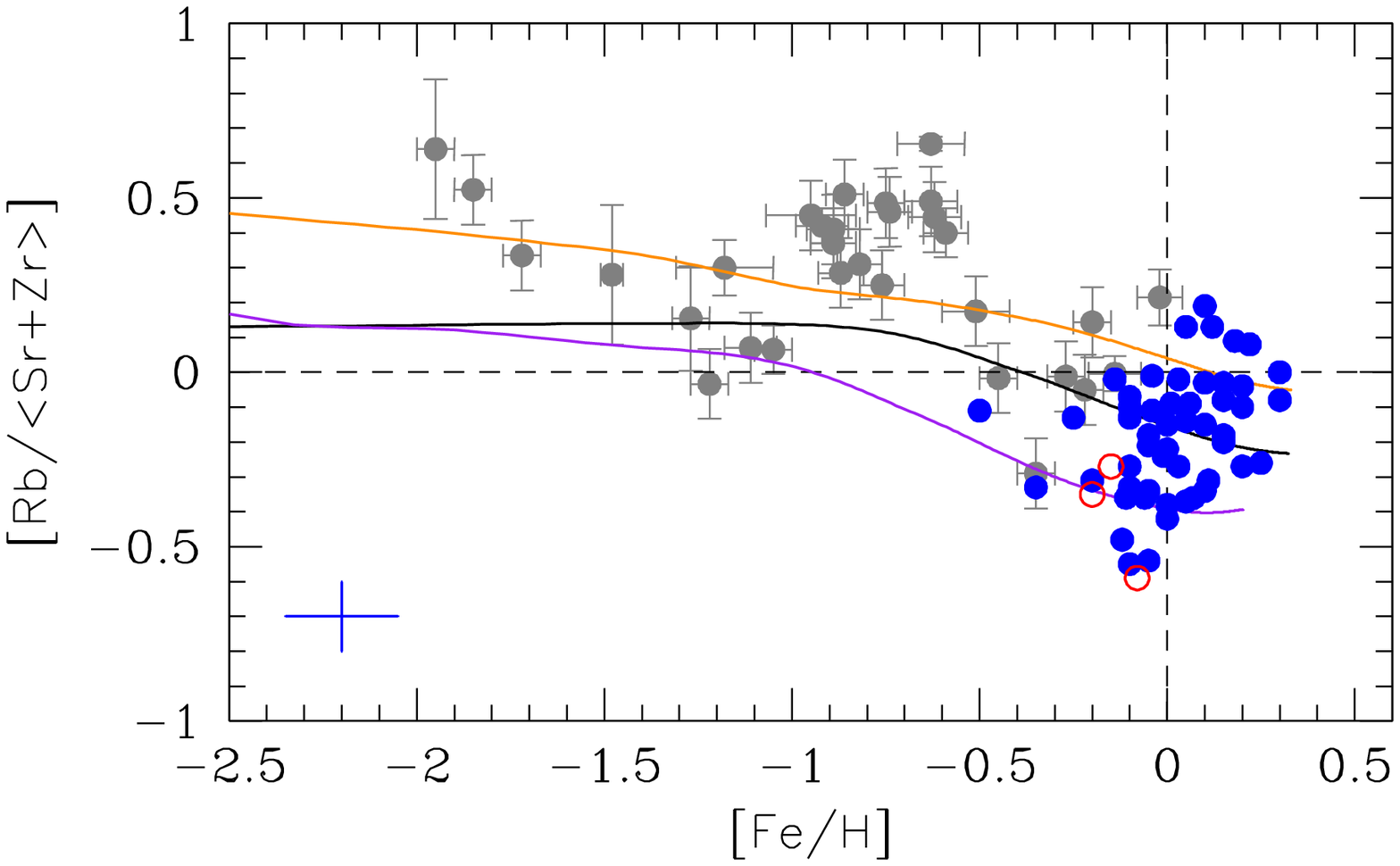}
   \caption{Same as Fig.~\ref{fig:[X/Fe]} but for [Rb/<Sr+Zr>] vs. [Fe/H]. Blue dots are M dwarfs in this study.
     Grey dots corresponds to the giants and dwarfs stars analysed in common by \citet{gra94}, \citet{tom99},
     and \citet{mis19}. Red open circles are stars in which Rb lines may be affected by the magnetic field.
     Coloured solid lines are the GCE predictions as in Fig. 5. The black solid line shows the GCE prediction
     when the weak s-process is inhibited in massive stars (i.e. no rotation).}
    \label{fig:[Rb/X]}
\end{figure}

As mentioned before, for metallicities above solar, we found an increasing [Rb/Fe] trend, which would imply a
secondary behaviour with metallicity. 
This trend is surprising, because according to GCE models, the Fe contribution from type Ia supernovae should
nominally decrease the trend. The secondary nature of this trend excludes a standard r-process contribution, while
it could favour a s-process scenario. The Rb s-process production is indeed secondary in both non-rotating massive
stars and AGB stars. This production is related to the activation of the $^{22}$Ne$(\alpha,n)^{25}$Mg neutron source,
whose efficiency increases with metallicity  because of the larger amount of $^{22}$Ne left by the burning of
CNO \footnote{At the end of the H burning, the initial amount of C+N+O is mostly transformed into $^{14}$N. Later on,
  when the He burning starts, $^{22}$Ne is produced by the chain $^{14}$N$(\alpha,\gamma)^{18}$F$(\beta)^{18}$O$(\alpha,\gamma)^{22}$Ne. In
  practice, the initial amount of C+N+O is transformed into $^{22}$Ne. This scenario is altered if rotation-induced mixing
  brings fresh C, as synthesised by the He burning, into the H-burning shell.}. 
In AGB stars in particular, this neutron source is active during the thermal pulses when the temperature in  the He-shell
attains or exceeds $\sim 300$ MK. However, in low-mass AGB stars, which are the galactic polluters responsible for the p
roduction of the s-process main component, the $^{22}$Ne$(\alpha,n)^{25}$Mg is only marginally activated because of the
low temperature developed during the thermal pulses. In contrast, the main neutron source, namely $^{13}$C$(\alpha,n)^{16}$O,
is active during the interpulse phase, when the temperature is  $\sim 90$ MK, but the resulting neutron density is much
lower than the threshold needed to open the $^{85}$Kr (ground state) and $^{86}$Rb branches. This occurrence is illustrated in  Figure 7. As a
result, low [Rb/Sr,Y,Zr] ratios are expected. 

An interesting deviation from the standard AGB scenario illustrated so-far was firstly described by \citet{cristallo2009}
\citep[see also][]{kar10,cristallo2018}.  At nearly solar metallicity, 
the $^{13}$C is not completely burnt during the interpulse and part of it is engulfed in the convective shell generated by
the following thermal pulse and burned at higher temperature ($\sim 200$ MK). The resulting neutron density is high enough
to open the $^{85}$Kr and $^{86}$Rb branches, allowing the production of the neutron magic rubidium isotope: $^{87}$Rb. In
this case, owing to the low neutron-capture cross section of this magic nucleus, an overproduction of Rb is expected,
while Sr, Y, and Zr, which are mainly produced by the interpulse $^{13}$C burning, are not affected. The occurrence of
this additional $^{13}$C neutron burst requires particularly low interpulse temperatures, and in turn low core masses. For
this reason, it may only occur in AGB stars with low masses (M$\lesssim 3$ M$_\odot$) and high metallicities ([Fe/H]$>-0.1$). An
important ingredient of the theoretical stellar models that hampers a definite validation of this hypothesis is our
poor knowledge of the $^{13}$C$(\alpha,n)^{16}$O reaction \citep{cristallo2018}. In this context, a lower rate at $T\sim 90$ MK
would greatly enhance the contribution to the galactic Rb of low-mass AGB stars at nearly solar and super-solar metallicity.
It is possible that the observed high-metallicity trend of galactic Rb shown in Fig. 5 (top panel) is a hint in favour of this scenario.

A similar trend of increasing [Ba/Fe] with increasing metallicity was reported at young ages in solar twins, open clusters,
and local associations \citep[e.g.][]{dor12,red15,nis16,tab17}. 
However, this trend is not seen for other heavy elements of similar s-process origin (e.g. La, Nd). 
To account for this Ba trend, \citet{mai12} suggested that the observed enhancement can be produced by nucleosynthesis
in AGB stars of low mass (M $< 1.5$\,M$_\odot$) if they release neutrons from the $^{13}$C$(\alpha,$n$)^{16}$O reaction in
reservoirs larger by a factor of four than assumed in more massive AGB stars (M $> 1.5$\,M$_\odot$).
\citet{mis15} instead reintroduced the intermediate neutron-capture i-process discussed earlier by  \citet{cow77} to
explain this fact, in which the neutron density operating the process is approximately $10^{15}$\,cm$^{-3}$ . 
Also, recently \citep{ryd20} found a secondary behaviour of [F/Fe] for supersolar metallicities in K giants. 
We note that F has a significant cosmic contribution from AGB stars \citep{jor92,kar14,abi19}. 
Although F is produced in AGB stars from an alternative nuclear channel to the s-elements, its production needs neutrons
and  is therefore linked to the s-process \citep[see e.g.][]{cri14,abi19}. 
Nevertheless, as far as the [Ba/Fe] ratio is concerned this trend has recently been shown to be correlated with the
stellar activity of young stars and to not be nucleosynthetic in origin \citep{red17}. 
Some of the stars studied here were flagged as active by \citet{pass19}, but except for the three stars mentioned above,
we checked that the rest of the stars flagged as active are distributed uniformly within the range of [Rb/Fe] ratios
derived here. In any case, a detailed study of possible connection between stellar activity and Rb abundances is
suggested in M dwarfs without showing any peculiar abundance pattern.

Another valuable piece of information on the production sources of Rb and its evolution with metallicity comes
from its abundance ratio with respect to the neighbouring s-elements Sr, Y, and Zr. 
As mentioned in Section 1 this abundance ratio provides relevant information on the physical conditions prevailing
at the s-process site, in particular about the neutron density. 
This may allow, for instance,  the characteristic stellar mass of the producing objects to be inferred. 
Figure~\ref{fig:[Rb/X]} shows the [Rb/<Sr+Zr>]-versus-[Fe/H]  ratios derived in our stars (blue dots), as well
as in the studies by \citet{gra94}, \citet{tom99}, and \citet{mis19} in  FGK dwarfs (grey dots). 
<Sr+Zr> is the average ratio between [Sr/H] and [Zr/H] in the corresponding star. 
As in Fig.~\ref{fig:[X/Fe]}, the orange line is the theoretical prediction according to the GCE model by \citet{pra18}, which
includes LIMS, MRS, and the r-process contributions, while the magenta line is the prediction when only yields from LIMS are included. 
Figure~\ref{fig:[Rb/X]} also shows the prediction in the case where rotation is excluded in massive stars (black line), that is,
the weak s-process contribution is strongly inhibited. 
We note that the observed trend does not change significantly if non-LTE Rb abundances are used. 
At low metallicity ([Fe/H] $<-1.0$) the observed  [Rb/<Sr+Zr>] ratio is above the solar value and again is better
reproduced when all the possible production sources of Rb are included (orange line). 
At these metallicities, the r-process dominates the production of Rb, Sr, and Zr but, as mentioned before, there is
a non-negligible contribution from the weak s-process in massive stars as can be appreciated by comparison with the
prediction with no weak s-process (black line). 
At the typical neutron densities of the weak s-process ($N_n \sim 10^{11-13}$\,cm$^{-3}$), some branchings, closed
in the case of the $^{13}$C$(\alpha,$n$)^{16}$O neutron source, are opened (see Fig. 7). This  allows  alternative
paths, in particular that at the $^{85}$Kr, which leads to an overproduction of Rb with respect to Sr and Zr,
giving [Rb/<Sr+Zr>]$\geq 0.0$. We note that the
$^{22}$Ne$(\alpha,$n$)^{25}$Mg source becomes relevant even for stellar masses as low as $\sim 2.5$\,M$_\odot$
and Z$\lesssim 0.01$\,Z$_\odot$ \citep{str14}. 
Figure~\ref{fig:[Rb/X]} shows that as metallicity increases, the [Rb/<Sr+Zr>] diminishes and is below solar
on average for [Fe/H]$\sim 0.0$, although with a significant dispersion.
This dispersion is considerably larger than expected from the uncertainties in the analysis and  may therefore be real. 
The decrease in the  [Rb/<Sr+Zr>] ratio for increasing metallicity is clearly due to the increasing relevance
of the contribution of low-mass stars in the production of Rb, Sr, and Zr through the main s-process, for which the
$^{13}$C$(\alpha,$n$)^{16}$O is the main neutron source.
When this neutron source is at work, [Rb/<Sr+Zr>]$<0.0$ is expected, as shown by GCE model (magenta line).
Nevertheless, in Fig. 8 there are a number of stars showing [Rb/<Sr+Zr>] $\gtrsim 0.0$. 
If the Rb abundances in these stars were correct (see the discussion above), these ratios would be better
explained if massive (intermediate) AGB stars contribute significantly at [Fe/H]$\gtrsim 0.0$.  
In summary, with the available yields for Rb from different sources, the observed [Rb/<Sr+Zr>] ratios at
near-solar metallicity seem to require a mix of these contributing sources.
It is obvious that determinations of Rb abundances in FGK dwarf stars, the analysis of which is subject
to less uncertainties, are needed to shed light on this Rb puzzle.

\section{Summary}
\label{summary}

We derived, for the first time, abundances of Rb and its associated elements Sr and Zr in a sample of
M dwarfs in the close solar neighbourhood and in the metallicity range $-0.5\lesssim$ [Fe/H] $\lesssim +0.3$. 
These metallicity and spectral type ranges were poorly explored in previous studies for Rb abundances. 
We used high-resolution and high-S/N spectra in the VIS and NIR acquired with the CARMENES spectrograph. 
The main novel result of our study is that, in the  explored metalliciy range and relative to the metallicity,
rubidium  is systematically underabundant with respect to the Sun. 
This underabundance is of almost a factor two on average, although with significant dispersion. 
This result is in contrast with the figure found for Sr and Zr ratios with respect to the metallicity,
which are very close to the solar ratio, and is identical to that found in unevolved FGK dwarfs of similar metallicity. 
Furthermore, we find a possible positive correlation of the [Rb/Fe] ratio with increasing metallicity. 
We discuss the 
reliability of these results for Rb, never found for other
  neutron-capture element in the metallicity range studied here,
in terms of non-LTE effects, systematic errors in the analysis, anomalous Solar System Rb abundance, and stellar activity,
but no plausible explanation is found. 

Under the assumption that the abundance trend is real, we interpret the [Rb/Fe]-versus-[Fe/H] relationship in the
observed full range of metallicity by comparing it with theoretical predictions from  state-of-the-art one-dimensional GCE
models for neutron-capture elements in the solar neighbourhood \citep{pra18}. The observed Rb overabundances at [Fe/H]
$\lesssim -1.0$ can be explained if the r-process is the main source of the Galactic Rb budget at these metallicities,
although with a significant contribution from the weak s-process
occurring in rotating massive stars. 
However, the [Rb/Fe] underabundances observed at higher metallicities cannot be reproduced by this model, nor can the
possible secondary behaviour of [Rb/Fe] with metallicity. This secondary behaviour would require a much higher Rb
yield from AGB stars through the main s-process than is currently obtained in s-process nucleosynthesis calculations
without overproducing Sr, Y,  and Zr. A possible nucleosynthesis scenario 
occurring in low-mass and high-metallicity AGB stars is suggested.
In addition, the negative [Rb/<Sr+Zr>] ratios observed 
at near solar and higher metallicities, when compared with the GCE predictions, indicate that AGB stars are indeed
the main producers of Rb at these metallicities, although the large scatter observed suggests that there are  other
contributing sources. 

Overall, our abundance analysis showcases the value of abundance determinations in M dwarfs  for Galactic chemical
evolution studies. Additional Rb abundance measurements in FGK dwarfs of near solar metallicity, as well as an
evaluation of the impact of stellar activity on  abundance determinations in M dwarfs, are urgently needed to confirm
or disprove the main findings of this study.

\begin{table*}
\caption{Stellar parameters and abundances derived in the sample of M dwarfs$^a$.}         
\label{par_ab_data}      
\centering          
\begin{tabular}{llccccccc}     
\hline
\hline       
\noalign{\smallskip}
Karmn & Name & $T_{\rm eff}$ (K) & $\log{g}$ & C/O & [M/H] & $\log{\epsilon({\rm Rb})}$  & $\log{\epsilon({\rm Sr})}$ & $\log{\epsilon({\rm Zr})}$ \\ 
\hline                    
\noalign{\smallskip}
 J00051$+$457 & \object{GJ~2}         &3772  & 4.68 & 0.59 & $0.10$ & $2.32\pm0.02$ & 2.91&2.70\\
J00183$+$440 & \object{GX~And}   & 3606 & 4.77 & 0.61 & $-0.14$ &  $2.12\pm0.05$ & 2.58&2.50\\
J00389$+$306 & \object{G~60--10} & 3491 & 4.78 & 0.59 & $-$0.10 &  $2.22\pm0.01$ & 2.73&2.55\\
J00570$+$450 & \object{G~172--30}& 3397 & 4.85 & 0.50 & $-$0.10 &  $2.13\pm0.01$ & 2.90&2.80\\
J01013$+$613 & \object{GJ~47}         &3496 & 4.80 & ... & $-$0.11 &  $2.00\pm0.02$ & 2.80&$2.55\pm0.02$\\
J01025$+$716 & \object{GJ~48}         &3432 & 4.78 & 0.47 & 0.00 & $2.19\pm0.01$ & 2.85&2.60\\
J01026$+$623 & \object{BD$+$61 195}&3810 & 4.67 & 0.54 & 0.06 & 2.45&  2.98&2.63\\
J01433$+$043 & \object{G~3--14}    &   3485 & 4.80 & 0.50 & $-$0.10 &  $1.99\pm0.03$ & 2.70&$2.60\pm0.01$\\
J01518$+$644 & \object{G~244--37}  & 3666 & 4.64 & 0.58 & 0.20 & $2.42\pm0.02$ & 2.96&$2.72\pm0.02$\\
J02015$+$637 & \object{G~244--47}           &3452 & 4.79 & 0.59 & 0.03 & $2.44\pm0.01$ & 2.90&$2.61\pm0.01$\\
J02123$+$035 & \object{BD$+$02 348}&3605 & 4.90 & 0.58 & $-$0.35 &  $1.80\pm0.02$ & 2.57&$2.45\pm0.03$\\
J02222$+$478 & \object{BD$+$47 612}           &3916 & 4.64 & 0.58 & 0.15 & $2.43\pm0.02$ & 2.95&$2.70\pm0.01$\\
J02358$+$202 & \object{BD$+$19 381}&3726 & 4.63 & 0.59 & 0.18 & $2.48\pm0.03$ & 2.93&$2.60\pm0.01$\\
J02442$+$255 & \object{G~36--31}    &3434 & 4.83 & 0.58 & 0.00 & $2.12\pm0.02$ & 2.98&$2.63\pm0.01$\\
J02565$+$554W &\object{G~174--19}   &3889 & 4.64 & 0.54 & 0.30 & $2.54\pm0.02$ & 2.98&$2.73\pm0.01$\\
J03181$+$382 & \object{BD$+$37 748} &3854 & 4.64 & 0.58 & 0.20 & $2.72\pm0.01$ & 3.20&$2.83\pm0.02$\\
J03213$+$799 & \object{GJ~133}           &3533 & 4.78 & 0.59 & $-$0.10 &  $2.05\pm0.05$ & ...& $2.50\pm0.01$ \\  
J03217$-$066 & \object{G~77--46}           &3502 & 4.85 & 0.54 & $-$0.10 &  $1.99\pm0.04$ & 2.98&$2.65\pm0.02$\\
J03463$+$262 & \object{HD~23453}   &4001 & 4.66 & 0.54 & 0.20 & $2.40\pm0.05$ & 3.11&$2.78\pm0.02$\\
J03531$+$625 & \object{Ross~567}   &3454 & 4.79 & 0.59 & $-$0.05 &  $2.00\pm0.02$ & 2.65&$2.48\pm0.01$\\
J04225$+$105  & \object{LSPM~J0422$+$1031}           &3415 & 4.80 & ... & 0.05 & $2.55\pm0.01$ & 2.89&2.73\\
J04290$+$219 & \object{HD~28343}   & 4187 & 4.63 & 0.54 & 0.30 & $2.65\pm0.01$ & 3.17&$2.85\pm0.05$\\
J04376$-$110 & \object{BD$-$11 916} &3611 & 4.74 & 0.59 & 0.05 & $2.12\pm0.05$ & 2.93&$2.70\pm0.01$\\
J04376$+$528 & \object{HD~232979}  &4010 & 4.64 & 0.58 & 0.00 & $2.10\pm0.05$ & 2.92&$2.63\pm0.01$\\
J04429$+$189 & \object{HD~285968}   &3689 & 4.66 & 0.58 & 0.12 & $2.54\pm0.01$ & 2.85&$2.60\pm0.01$\\
J04538$-$177 & \object{L~736-30}   &3543 & 4.76 & 0.58 & $-$0.10 &  $2.06\pm0.01$ & 2.60&$2.50\pm0.01$\\
J04588$+$498 & \object{BD$+$49 1280}&4004 & 4.64 & 0.52 & 0.15 & $2.36\pm0.03$ & 3.00&$2.79\pm0.04$ \\
J05127$+$196 & \object{GJ~192}      &3570 & 4.75 & 0.54 & $-$0.01 &  $2.19\pm0.02$ & 2.87&$2.78\pm0.02$\\
J05314$-$036 & \object{HD~36395}    &3700 & 4.95 & 0.67 & 0.25 & $2.70\pm0.02$ & 3.40&$2.97\pm0.02$ \\  
J05365$+$113 & \object{HD~245409}   &4062 & 4.66 & 0.43 & 0.05 & $2.24\pm0.03$ & 3.05&$2.79\pm0.01$\\
J05415$+$534 & \object{HD~233153}   &3869 & 4.63 & 0.63 & 0.15 & $2.38\pm0.02$ & 3.00&$2.82\pm0.02$\\
J06103$+$821 & \object{GJ~226}            &3500 & 4.90 & 0.57 & $-$0.05 &  $2.07\pm0.02$ & 2.85&$2.65\pm0.05$\\
J06105$-$218 & \object{HD~42581}    &3842 & 4.68 & 0.54 & 0.10 & $2.43\pm0.02$ & 2.90&$2.66\pm0.03$\\
J06371$+$175 & \object{Ross~85}            &3732 & 4.77 & 0.63 & $-$0.25 &  $1.83\pm0.02$ & 2.40&2.37\\
J06548$+$332 & \object{Wolf~294}    &3410 & 4.85 & 0.57 & 0.00 & $2.28\pm0.05$ & 2.87&$2.62\pm0.01$\\
J07044$+$682 & \object{GJ~258}            &3421 & 4.81 & 0.57 & 0.05 & $2.32\pm0.01$ & 2.90&$2.65\pm0.01$\\
J07287$-$032 & \object{GJ~1097}            &3420 & 4.81 & 0.58 & 0.01 & $2.30\pm0.02$ & 2.83&$2.63\pm0.02$\\
J07353$+$548 & \object{GJ~3452}            &3513 & 4.81 & 0.59 & $-$0.05 &  $2.08\pm0.01$ & 2.70&$2.63\pm0.01$\\
J07361$-$031 & \object{BD$-$02 2198}& 3790 & 4.70 & ... & $-$0.04 &  $2.05\pm0.02$ & 2.60&$2.59\pm0.02$\\
J08161$+$013 & \object{G~113--0}    &3534 & 4.77 & 0.57 & $-$0.04 &  $2.15\pm0.01$ & 2.60&$2.60\pm0.01$\\
J08293$+$039 & \object{2M~J08292191+0355092}            &3685  & 4.64 & 0.54 & 0.22 & $2.52\pm0.01$ & 2.88&$2.63\pm0.01$\\
J08358$+$680 & \object{G~234--37}            &3434 & 4.82 & 0.57 & $-$0.06 &  $2.05\pm0.01$ & 2.85&$2.64\pm0.03$\\
J09133$+$688 & \object{G~234--57}             &3494 & 4.86 & ... & $-$0.05 &  $2.00\pm0.05$ & 2.98&$2.70\pm0.05$\\
J09140$+$196 & \object{LP~427--16}   &3591 & 4.65 & 0.58 & 0.11 & $2.20\pm0.05$ & 2.95&$2.59\pm0.01$\\
J09143$+$526 & \object{HD~79210}    &4007 & 4.67 & 0.58 & 0.07 & $2.17\pm0.02$ & 2.97&$2.68\pm0.01$ \\  
J09144$+$526 & \object{HD 79211}    & 3980 & 4.67 & 0.49 & 0.10 & $2.17\pm0.01$ & 2.95&$2.71\pm0.02$\\
J09163$-$186 & \object{L~749--34}    &3524 & 4.85 & 0.57 & $-$0.12 &  $1.97\pm0.01$ & 2.89&$2.72\pm0.02$ \\  
J09411$+$132 & \object{Ross~85}            &3654 & 4.72 & 0.57 & 0.03 & $2.19\pm0.02$ &  2.90&$2.65\pm0.01$\\
J09425$+$700 & \object{GJ~360}            &3718 & 4.67 & 0.41 & 0.10 & $2.55\pm0.02$ & 2.80&2.80\\
J11054$+$435 & \object{BD$+$44 2051A}& 3617 & 4.79 & 0.43 & $-$0.20 &  $2.05\pm0.02$ & 2.80&2.53\\
J11201$-$104\tablefootmark{b}  & \object{LP~733--099}            &3469 & 4.91 & 0.58 & $-$0.20 &  $1.78\pm0.02$ & 2.57&$2.50\pm0.01$\\
J12111$-$199 & \object{LTT~4562}    &3440 & 4.83 & 0.59 & $-$0.10 &  $2.18\pm0.02$ & 2.75&2.70\\
J13450$+$176  & \object{BD$+$18 2776}&3900 & 4.80 & 0.57 & $-$0.50 &  $1.75\pm0.05$ & 2.30&2.20\\
J13457$+$148 & \object{HD~119850}   & 3610 & 4.73 & 0.47 & $-$0.10 &  $2.12\pm0.02$ & 2.58&2.65\\
J15218$+$209\tablefootmark{b} & \object{OT~Ser}            &3500 & 4.90 & 0.58 & 0.10&2.05 &    2.71&$2.70\pm0.02$\\
J18174$+$483\tablefootmark{b}  & \object{TYC~3529-1437-1}      &3435 & 4.80 & 0.59 & $-$0.08 &  $1.85\pm0.02$&2.88&$2.58\pm0.01$\\
J22565$+$165 & \object{HD~216899}   &3606 & 4.77 & 0.59 & 0.15 & $2.55\pm0.01$ &   3.02&2.75 \\    
\noalign{\smallskip}
\hline                  
\end{tabular}
\tablefoot{$^{(a)}$ The C/O ratio in the Sun is 0.575 \citep{lod19}. 
Abundances of Rb, Sr, and Zr are given on the scale $\log{N({\rm H})} \equiv 12$. 
The Sr abundances shown are corrected for non-LTE effects (see text). For Rb and Zr abundances, the abundance
dispersion is also tabulated. $^{(b)}$ The resonance Rb lines in these stars may be affected by magnetic field.
Their abundance ratios are indicated as red open circles in Figs. 5 and 7.}
\end{table*}

\begin{acknowledgements}
  CARMENES is an instrument for the Centro Astron\'omico Hispano-Alem\'an (CAHA) at Calar Alto (Almer\'{\i}a, Spain),
  operated jointly by the Junta de Andaluc\'ia and the Instituto de Astrof\'isica de Andaluc\'ia (CSIC).
CARMENES was funded by the Max-Planck-Gesellschaft (MPG), 
  the Consejo Superior de Investigaciones Cient\'{\i}ficas (CSIC),
  the Ministerio de Econom\'ia y Competitividad (MINECO) and the European Regional Development Fund (ERDF) through
  projects FICTS-2011-02, ICTS-2017-07-CAHA-4, and CAHA16-CE-3978, 
  and the members of the CARMENES Consortium 
  (Max-Planck-Institut f\"ur Astronomie,
  Instituto de Astrof\'{\i}sica de Andaluc\'{\i}a,
  Landessternwarte K\"onigstuhl,
  Institut de Ci\`encies de l'Espai,
  Institut f\"ur Astrophysik G\"ottingen,
  Universidad Complutense de Madrid,
  Th\"uringer Landessternwarte Tautenburg,
  Instituto de Astrof\'{\i}sica de Canarias,
  Hamburger Sternwarte,
  Centro de Astrobiolog\'{\i}a and
  Centro Astron\'omico Hispano-Alem\'an), 
  with additional contributions by the MINECO, 
  the Deutsche Forschungsgemeinschaft through the Major Research Instrumentation Programme and Research Unit FOR2544 ``Blue Planets around Red Stars'', 
  the Klaus Tschira Stiftung, 
  the states of Baden-W\"urttemberg and Niedersachsen, 
  and by the Junta de Andaluc\'{\i}a.
  We acknowledge financial support from the Agencia Estatal de Investigaci\'on of the Ministerio de Ciencia e Innovaci\'on,
  the Universidad Complutense de Madrid, the Funda\c{c}\~{a}o para a Ci\^{e}ncia e a Tecnologia, the Generalitat de Catalunya,
  ERDF, and NASA through projects 
  PGC2018-095317-B-C21, 
  PID2019-109522GB-C51/2/3/4,   
  PGC2018-098153-B-C33,         
  AYA2016-79425-C3-[1,2,3]-P, 
  FPU15/01476, 
  UID[B,P]/04434/2020, 
  PTDC/FIS-AST/28953/2017, 
  POCI-01-0145-FEDER-028953, 
  SEV-2015-0548, 
  SEV-2017-0709, 
  MDM-2017-0737, 
  NNX17AG24G, 
  and the CERCA programme.
Finally, we thank to Verne Smith, Katia Cunha, and the
APOGEE/ASPCAP team for the APOGEE line list, and to K. Lodders for her
useful comments on the manuscript.

\end{acknowledgements}
\bibliographystyle{aa}
\bibliography{39032corr-bis}

\end{document}